\documentclass[12pt,titlepage]{article}
\usepackage{tabularx}
\usepackage{aliascnt}
\usepackage{amscd}
\usepackage{amsfonts}
\usepackage{amsmath}
\usepackage{amssymb}
\usepackage{bbm}
\usepackage{amsthm}
\usepackage{array}
\usepackage[english]{babel}
\usepackage{dsfont}
\usepackage{enumerate}
\usepackage{graphicx}
\usepackage{caption}
\usepackage{subcaption}
\usepackage{hyperref}
\usepackage{latexsym}
\usepackage{multirow}
\usepackage{epstopdf}
\usepackage{mathrsfs}
\usepackage{natbib}
\usepackage{pdfsync}
\usepackage{pgfplots}
\usepackage{epstopdf}
\usepackage{tikz}
\usepackage{algorithm}
\usepackage{algpseudocode}
\usepackage{color}
\usepackage{graphicx,psfrag,epsf}
\usepackage{enumerate}
\usepackage{url}
\usepackage{amsthm,bm}

\newaliascnt{rem}{thm}

\aliascntresetthe{rem}
\renewenvironment{proof}{{\em Proof.} }{\hfill $\Box$}
\usepackage{tabu}
\usepackage{makecell}

\def\1{\mathds{1}}

\usepackage{geometry}
\begin{document}
\pagenumbering{roman}
\title{Probabilistic Forecasting of the Arctic Sea Ice Edge with Contour Modeling \thanks{ The authors thank Nicholas Wayand for assistance with data processing and Arlan Dirkson for addressing questions on Trend Adjusted Quantile Mapping. Any opinions, findings, and conclusions or recommendations expressed in this material are those of the authors and do not necessarily reflect the views of the National Science Foundation.  Results in this paper were produced using the \textbf{\it{IceCast}} R package  \citep{Director2020}. Relevant code can also be accessed at \url{https://github.com/hdirector/ProbSeaIce. }}}

\author{Hannah M. Director\\ Department of Statistics \\ University of Washington \and
Adrian E. Raftery \\ Departments of Statistics and Sociology \\University of Washington \and
Cecilia M. Bitz \\ Department of Atmospheric Sciences \\ University of Washington 
}

\maketitle

\begin{abstract}
Sea ice, or frozen ocean water, freezes and melts every year in the Arctic. Forecasts of where sea ice will be located weeks to months in advance have become more important as the amount of sea ice declines due to climate change, for maritime planning and other uses.  Typical sea ice forecasts are made with  ensemble models, physics-based  models of sea ice and the surrounding ocean and atmosphere. This paper introduces Mixture Contour Forecasting, a method to forecast sea ice probabilistically using a mixture of two distributions, one based on post-processed output from ensembles and the other on observed sea ice patterns in recent years. At short lead times, these forecasts are better calibrated than unadjusted dynamic ensemble forecasts and other statistical reference forecasts.  To produce these forecasts, a statistical technique is introduced that directly models the sea ice edge contour, the boundary around the region that is ice-covered.  
Mixture Contour Forecasting and reference methods are evaluated for monthly sea ice forecasts for 2008-2016 at lead times ranging from 0.5-6.5 months using one of the European Centre for Medium-Range Weather Forecasts ensembles. \\

\end{abstract}

\begin{small}
\newpage
\tableofcontents

\newpage

\listoftables

\medskip

\listoffigures
\end{small}

\newpage
\pagenumbering{arabic}
\baselineskip=18pt

\section{Introduction}
\label{sec:intro}
Sea ice, or frozen ocean water, freezes and melts annually in response to seasonal changes in atmospheric and oceanic processes. Since the satellite record began in 1979, the amount of sea ice in the Arctic has declined rapidly \citep{Comiso2008, Stroeve2012}. Continued reduction in sea ice is expected as the effects of climate changes increase. Reduced sea ice cover allows for increased Arctic shipping \citep{Smith2013, Melia2016}. The importance of forecasting sea ice  has increased in response, since waters without sea ice are more easily navigable than waters with sea ice. Reliable estimates of a ship's probability of encountering sea ice are needed to plan maritime routes that avoid sea ice.  In this paper, we develop statistical methods to accurately predict the probability of encountering sea ice. 

Sea ice concentration, or the percent of ice-covered area, has been derived from satellite measurements for a little over 40 years  and is reported on a grid.  For navigational purposes, the concentration field can be reduced to a binary field indicating the presence or absence of sea ice. Prediction efforts then focus on the location of the ice edge contour,  or the boundary line that separates ice-covered regions and open water. We follow the convention in sea ice research of categorizing a grid box as ice-covered if its concentration is at least 15$\%$. Thresholding is needed, since satellites often fail to distinguish between areas of open water and areas where water is melting on the sea ice's surface. Concentration reduces from about 50$\%$ to near 0$\%$ concentration over a small region, so the area classified as sea ice is only weakly affected by the exact threshold concentration used.  

\citet{Zhang2019, Zhang2020} have introduced hierarchical spatio-temporal generalized linear models for Arctic sea ice. However, many sea ice forecasts used in practice are informed by numerical prediction systems.  These systems integrate systems of differential equations to represent the physical processes that drive sea ice formation and melting. These systems are typically run multiple times with slightly different initial conditions, and the outputs from the resulting runs have varying amounts of sea ice. 
The collection of forecasts, referred to as the ensemble, has shown skill in predicting the total area or extent of sea ice at seasonal time scales in retrospective forecasts \citep[e.g., ][]{Sigmond2013, Msadek2014, Wang2013, Chevallier2013} and in current forecasts \citep{Blanchard2015}. Skill has also been shown at regional scales \citep{Bushuk2017} and for spatial fields for some models at short lead times \citep{Zampieri2018}. 

However, errors in ensembles are common because the underlying systems of differential equations are only approximations of the true physical processes, because initial conditions are not fully known, and because of sub-grid scale phenomena  \citep{Guemas2016, Blanchard2015}. 
An ensemble can be biased, meaning that its mean behavior is systematically incorrect.
It can also be poorly calibrated, meaning that the range of possible sea ice states predicted by the ensemble members does not reflect the actual uncertainty of the forecast. 

Statistical post-processing, or methods that incorporate or adjust information from ensemble forecasts, can be applied to address ensembles' weaknesses while maintaining much of the skill they provide. In this paper, we develop Mixture Contour Forecasting (MCF), a post-processing method to improve the calibration of sea ice forecasts. First, a method for generating distributions of sea ice edge contours is developed. The mean location of the sea ice edge contour in these distributions is partially informed by the mean location of the sea ice edge contour obtained from ensemble outputs. The forecasts obtained from these generated contour distributions are then weighted with climatological information to account for the time-varying skill of ensemble forecasts and aspects of sea ice that cannot be represented well with a contour boundary, such as holes in the sea ice. 

\begin{figure}[htb]
	\centering
	\includegraphics[width = 1.0\textwidth]{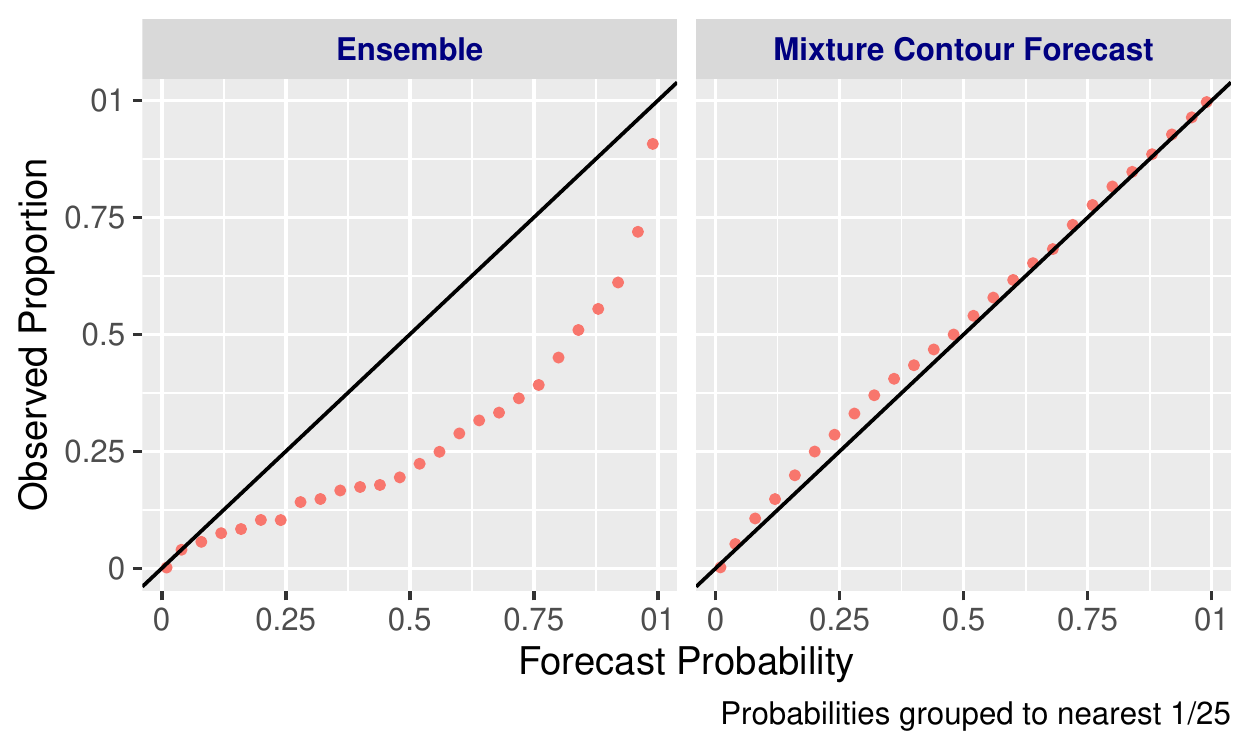} 
	\caption{The average proportion of times sea ice was present, plotted against the predicted probability of sea ice presence for the unadjusted ECMWF ensemble forecasts (left) and for the corresponding Mixture Contour Forecasts (right) for lead times of 0.5 to 1.5 months. Results are for September 2008-2016. A perfectly calibrated forecast would have all points on the y = x line, so the MCF method forecast is better calibrated. }
	\label{fig: calib_sep}
\end{figure}

The MCF method provides better calibrated and more accurate probabilistic forecasts than the unadjusted ensemble and better calibrated forecasts than existing post-processing techniques. In Figure \ref{fig: calib_sep}, we illustrate the extent to which MCF improves model calibration by plotting the predicted probability of sea ice presence in September against the actual proportion of times sea ice was observed for the raw ensemble and after post-processing. The predictions are from the fifth generation of the European Centre for Medium-Range Weather Forecasts (ECMWF) seasonal forecasting season (SEAS5) \citep{Johnson2019, ECMWF2017}.  We see that MCF provides much improved model calibration.

The paper is organized as follows.  In Section \ref{sec: contourModel}, we introduce a Bayesian model for generating distributions of the sea ice edge contour. This contour model is fit to observed ice edge contours from recent years and its prior is partially informed from the mean ice edge predicted from a dynamic ensemble. In Section \ref{sec: MCF}, the contour model is combined with climatological information using a finite mixture model. In Section \ref{sec: methodEvaluation}, we compare the performance of MCF to other post-processing and statistical forecasting techniques. In Section \ref{sec: discuss}, we conclude with discussion.

\section{Contour model}
\label{sec: contourModel}
In this section, we develop a Bayesian model for the distribution of sea ice edge contours. The method works by directly modeling contours as a sequence of connected points.

\subsection{Notation and setup}
\label{sec: notationAndSetup}
A contour is the boundary line enclosing a defined area, which in this case is the region that contains sea ice. A contour, denoted by $\boldsymbol{S}$, can  be represented as an ordered sequence of $N$ spatial points, $(S_{1}, \hdots, S_{N})$, where each $S_{i}$ is an $(x,y)$ coordinate pair. Connecting $S_{i}$ to $S_{i+1}$ for  all $i = 1, \hdots, N -1$ and $S_{N}$ to $S_{1}$ encloses an area.  Following this definition, to generate a distribution of contours we need a way to generate realizations of $\boldsymbol{S}$. 

\begin{figure}[!htb]
	\centering
	\includegraphics[height = .50\textheight]{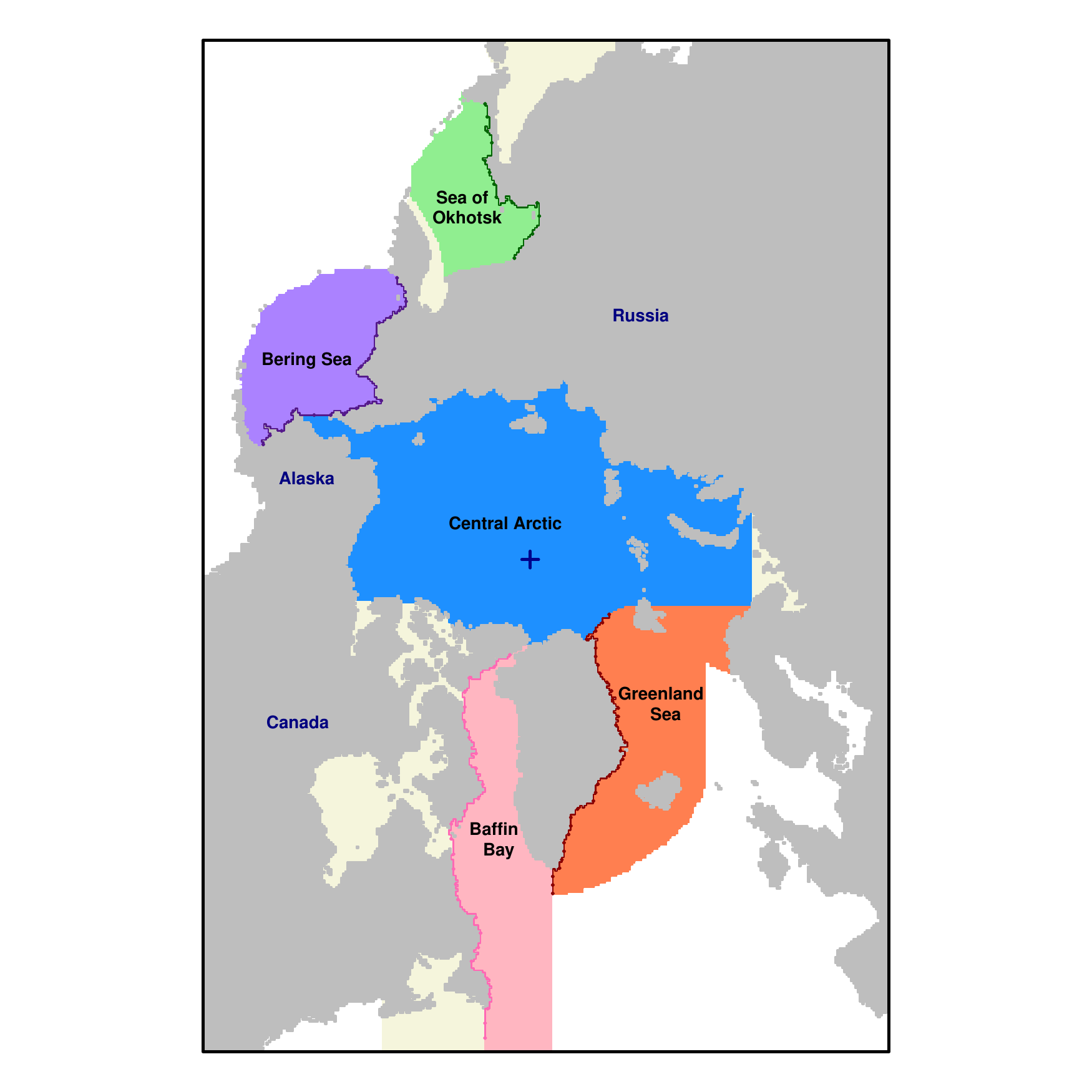} 
	\caption{Arctic ocean regions used here.  Each region in a color other than beige is fit with a contour model. The bold lines in all regions except the Central Arctic are the lines from which the fixed set of boundary points $\boldsymbol{B}$ are drawn. The boundary points themselves are plotted on top of these lines. The  `+' symbol denotes the location of $\boldsymbol{B}$ in the Central Arctic Region. Areas in grey are land and areas in white are ocean regions that are not considered part of the Arctic ocean in the National Snow and Ice Data Center land mask \citep{nsidc2016}. }
	\label{fig: regionMap}
\end{figure}

While the sea ice edge is often referred to as a single entity, it is actually a collection of edges defining multiple contiguous areas of sea ice. As such, it is natural to model multiple contours separately. We focus on five regions individually. These five regions, shown in the map in Figure \ref{fig: regionMap}, exclude parts of the Arctic ocean where a contour model is not appropriate because the sea ice does not typically form one contiguous section. We selected these regions by modifying an existing region mask  \citep{Cavalieri2012}  obtained from the \citet{nsidc2016}.    For notational simplicity, we do not subscript the regions and refer to the sea ice edge contour in a given region simply as $\boldsymbol{S}$.

In most regions, sea ice is formed in contiguous sections bordering land.  In these regions, $\boldsymbol{S}$ is formed by a sequence of points that proceed from the coastline, into the ocean, and back to the coastline. We can reduce the number of points that need to be estimated by fixing a set of boundary points, $\boldsymbol{B} = (B_{1}, \hdots, B_{n})$, on land and considering how far into the ocean the contour extends at each boundary location. The subset of points in $\boldsymbol{S}$ which must be fit are denoted by $\boldsymbol{\tilde{S}}$ and are indexed $\{1, 2, \hdots, n \}$.  We lay out an ordered series of parallel lines, $\boldsymbol{L} = (L_{1}, \hdots, L_{n})$, that cover the region. Each $L_{i}$ extends from its corresponding point $B_{i}$ to the edge of the region.  We assume that one point, $\tilde{S}_{i}$, lies on each line, $L_{i}$.  We denote the line segment from each point on the coastline, $B_{i}$, to the corresponding point $\tilde{S}_{i}$ as $y_{i}$.  The set of all line segments is denoted by $\boldsymbol{Y} = (y_{1}, \hdots, y_{n})$. The contour is then formed by connecting the points $B_{i}$ to $B_{i + 1}$ for all $i = 1, \hdots, n-1$, $B_{n}$ to $\tilde{S}_{n}$, $\tilde{S}_{i}$ to $\tilde{S}_{i-1}$ for all $i = n, \hdots, 2$, and $\tilde{S}_{1}$ to $B_{1}$. The left panel of Figure \ref{fig: boundLines} illustrates these values for the Bering Sea.  The angle of all lines in $\boldsymbol{L}$ is set to approximately match the direction the sea ice grows off the land in each region. 

Unlike in other regions, sea ice in the Central Arctic region is not generally formed off a land boundary. To represent the Central Arctic's  contour, we fix all the lines in $\boldsymbol{L}$ to originate from a single fixed central point rather than from a sequence of points. So, for this region $B_{i} = B_{j}$ for all $B_{i}, B_{j} \in \boldsymbol{B}$. The lines extend at fixed angles evenly spaced around a circle as illustrated in the right panel of Figure \ref{fig: boundLines}. In this case, $\boldsymbol{\tilde{S}} = \boldsymbol{S}$ and $n = N$.

\begin{figure}[!htb]
	\centering
	\includegraphics[width=  .9\textwidth]{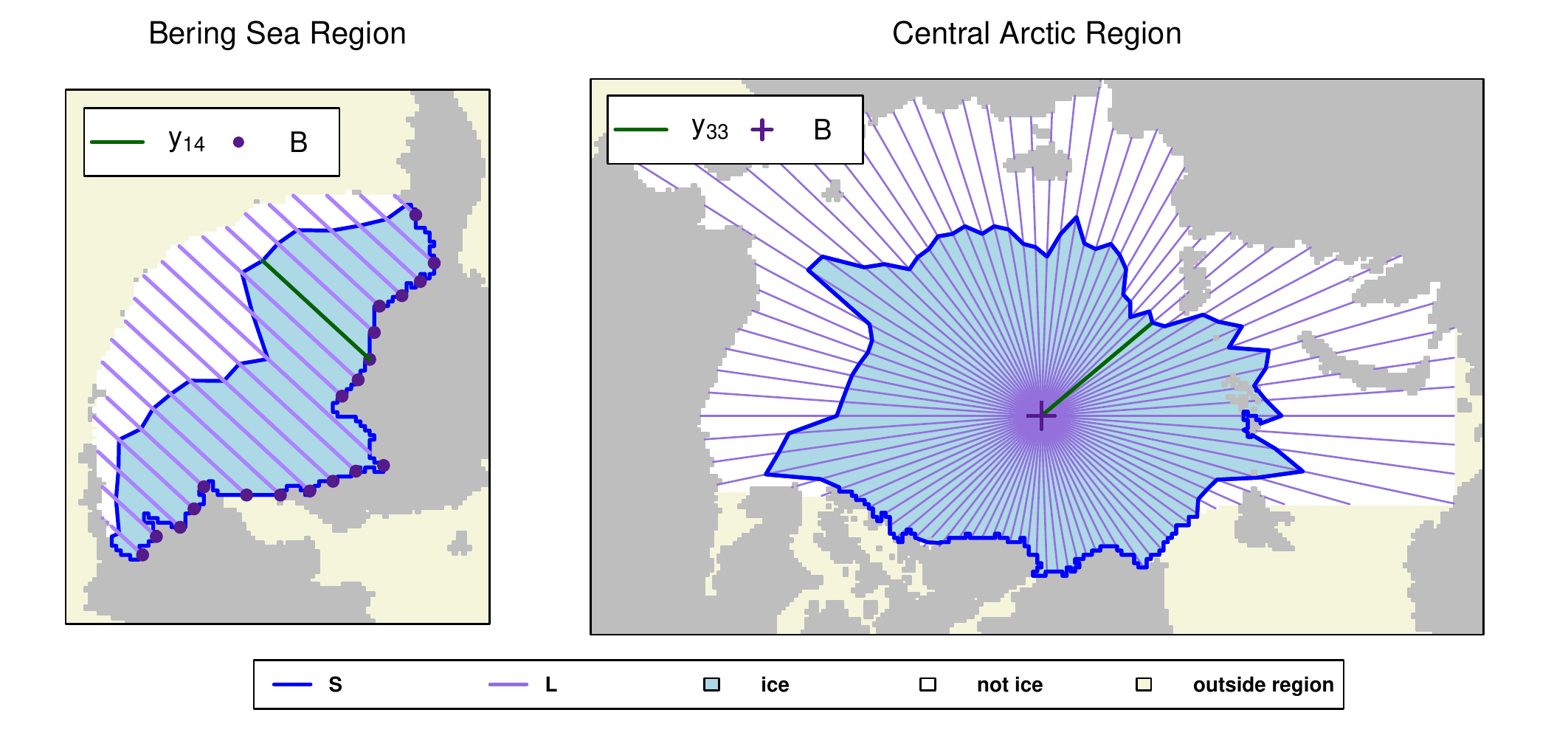} 
	\label{fig: boundLines}
	\caption{Hypothetical sea ice edge contours ($\boldsymbol{S}$), sets of fixed boundary points ($\boldsymbol{B}$), and  parallel lines ($\boldsymbol{L}$) on which the points $\boldsymbol{\tilde{S}}$ will be generated for a sample typical region (left) and for the Central Arctic region (right). The green line designates the observed ice-covered line segments for the 14th (left) and 33rd (right) lines.}
	\label{fig: boundLines}
\end{figure}

Note that given $\boldsymbol{B}$, $\boldsymbol{Y}$, and the angles of all lines in $\boldsymbol{L}$, we have enough information to identify each $\tilde{S}_{i}$. We need only compute the length of each line segment  $y_{i} \in \boldsymbol{Y}$. Each coordinate of the contour is then,
\begin{equation}
\label{eq: obviTrig}
\tilde{s}_{i} = \boldsymbol{B}_{i} + (||y_{i}||\cos (\theta_{i}), ||y_{i}|| \sin (\theta_{i}))
\end{equation}
where $||\cdot||$ denotes the length of line segment and $\theta_{i}$ is the angle of line $L_{i}$. Therefore, to generate distributions of contours, we need only develop a statistical model for generating the length of the line segments in $\boldsymbol{Y}$,  since $\boldsymbol{B}$ and $\boldsymbol{L}$ are fixed. 

\subsection{Statistical model}
For the sea ice application, each line $L_{i}$ is bounded below by zero and above by land and regions boundaries. Additionally, some $L_{i}$ cross over land sections, where sea ice cannot be observed.  These constraints make it natural to model the proportion of each line that is ice-covered, rather than model the length of the line segments that compose $\boldsymbol{Y}$ directly.

\begin{figure}[!htb]
\centering
	\includegraphics[width=  .9\textwidth]{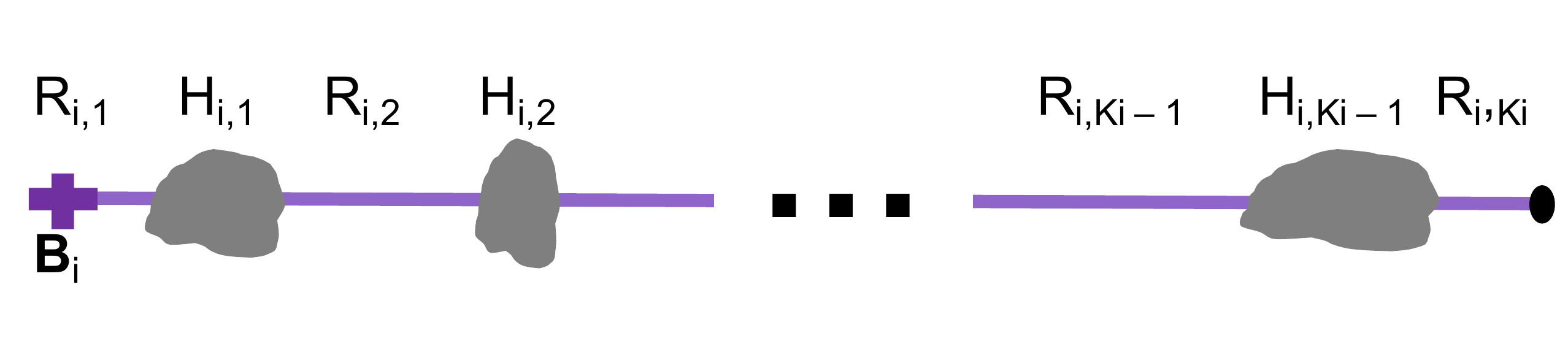} 
	\caption{Illustration of a hypothetical line segment, $L_{i}$ that  crosses over multiple sections of land. Line $L_{i}$ starts at point $B_{i}$, denoted by a `\textbf{\bf{+}}' sign, and ends at the black circle. The $j$-th line segment crossing ocean in $L_{i}$ is denoted by $R_{i,j}$ and the $j$-th line segment crossing land is denoted by $H_{i, j}$. There are $K_{i}$ sections crossing ocean and $K_{i} - 1$ sections crossing land.}
	\label{fig: propVars}
\end{figure}

 We now introduce notation for modeling proportions. These variables are illustrated in Figure  \ref{fig: propVars}.
Let $\boldsymbol{R}_{i} = \{R_{i,1}, \hdots, R_{i,K_{i}}\}$ denote the $K_{i}$ line segments that form  $L_{i}$ and let $\boldsymbol{H}_{i} = \{H_{i,1}, \hdots, H_{i,K_{i} -1}\}$ denote the $K_{i} - 1$ line segments where land is crossed. Note that  
\begin{equation}
||L_{i}|| = \sum_{k=1}^{K_{i}} ||R_{i,k}|| +  \sum_{k=1}^{K_{i} - 1} ||H_{i,k}|| = ||\boldsymbol{R}|| + ||\boldsymbol{H}||.
\end{equation}
  In the  common case where $L_{i}$ just goes through ocean, $\boldsymbol{R} = R_{i1} = L_{i}$ and $H = \emptyset$.

 Since the line segments forming any $H_{i}$ cannot contain sea ice, we focus on modeling the proportion of the corresponding $R_{i}$ that are ice-covered. More formally, let 
\begin{equation}
\pi_{i} = \frac{|| y_{i} \cap \boldsymbol{R}_{i}||
}{||\boldsymbol{R}_{i}||} ,
\end{equation}
where the numerator denotes the length of $y_{i}$ that intersects the ocean line segments and the dominator denotes the total length of of ocean line segments in $L_{i}$. The set of all proportions is denoted by $\boldsymbol{\pi} = (\pi_{1}, \hdots, \pi_{n})$.

We develop a model for $\boldsymbol{\pi}$ that can be used to generate $\boldsymbol{Y}$ and corresponding  $\boldsymbol{\tilde{S}}$.  For ease of modeling, we transform the proportions to the real line. Let
\begin{equation}
\tilde{\pi}_{i} = \begin{cases} \text{logit}(\pi_{i}) & \text{ for } \epsilon \leq \pi_{i} \leq 1 - \epsilon\\
\text{logit}(\epsilon) & \text{ for } \pi_{i}  < \epsilon\\
\text{logit}(1 - \epsilon) & \text{ for } \pi_{i} > 1 - \epsilon ,
\end{cases}
\end{equation}
where $\text{logit}(x) = \log(x/(1-x))$ and $\epsilon$ is small. In our implementation $\epsilon = 0.01$. The set of transformed proportions, $\boldsymbol{\tilde{\pi}}$, are modeled using a multivariate normal distribution,
\begin{equation}
\label{eq: piTildeNorm}
\boldsymbol{\tilde{\pi}}  \sim N(\boldsymbol{\mu}, \boldsymbol{\Sigma}).
\end{equation}
where $\boldsymbol{\mu}$ is an $n \times 1$ mean vector and $\boldsymbol{\Sigma}$ is an $n \times n$ covariance matrix.

The data generating process for $\boldsymbol{\tilde{S}}$ is then as follows. First an underlying random vector, $\boldsymbol{\tilde{\pi}}$, is drawn,  Then each $\tilde{\pi}_{i}$ is transformed back to a proportion via\begin{equation}
\pi_{i} = \text{ ilogit}(\tilde{\pi}_{i}) ,
\end{equation}
where $\text{ilogit}(x) = \exp(x)/(1 + \exp(x))$. 
The length of the  corresponding $y_{i}$ can be computed from $\pi_{i}$ as follows. Let $D$ denote the maximum index of a line segment of $R_{i}$ that is fully ice-covered when a proportion $\pi_{i}$ of $\boldsymbol{R}_{i}$ is ice-covered: 
\begin{equation}
D = \underset{d}{\text{argmin}}  \left\{ \frac{\sum_{k=1}^{d}||R_{i,k}||}{||\boldsymbol{R}_{i}||} < \pi_{i}\right\}.
\end{equation}
Then, the length of $y_{i}$ is  
\begin{equation}
||y_{i}|| = \pi_{i} ||\boldsymbol{R}_{i}|| + \sum_{k = 1}^{D - 1} ||H_{i,k}||.
\end{equation}
In other words, $y_{i}$ is composed of  a proportion $\pi_{i}$ of $\boldsymbol{R}_{i}$ and all the line segments of $\boldsymbol{H}_{i}$ that must be crossed to reach the $D$-th segment of $\boldsymbol{R}_{i}$. For all $i$, the lengths of $y_{i}$ are then used to compute $\boldsymbol{s}_{i}$ using Equation \ref{eq: obviTrig}. Connecting the points in $\boldsymbol{\tilde{S}}$, along with the points in $\boldsymbol{B}$, if applicable, produces a generated contour.

In rare cases, the generated values in $\boldsymbol{\tilde{S}}$ will result in a contour that intersects itself. When these self-intersections occur, the contour fails to be a boundary around a single contiguous area. In such cases, a small adjustment is made with the Douglas-Peuker algorithm to the part(s) of the contours that have self-intersections. The Douglas-Peucker algorithm takes as input a line represented as a connected sequence of points  and returns a new line that approximates the original line with a different connected sequence of points.  The new line uses as few points as possible to approximate the old line while ensuring that the new line differs from the old line by no more than a distance of $\eta$ \citep{Douglas1973}. To correct self-intersections, the Douglas-Peucker algorithm is initially applied with a small $\eta$ to the part of a contour with a self-intersection.  If the self-intersection is removed, the new line is used in place of the old line. If not, $\eta$ is increased and the algorithm is reapplied. This process is repeated until an $\eta$ is found that produces a contour with no self-intersections. In our context, these adjustments typically have minimal effect on the line itself and the area contained within the contour.

We also adjust a small percent of generated $||y_{i}||$ values that correspond to locations where the generated contour is very close to touching a region boundary or land (within 12.5 nominal kilometers, or half the nominal length of a grid box). In such cases, our statistical model is predicting that the contour comes very close to, but does not actually touch, a land or region boundary.  This behavior is physically implausible near land, so we adjust these $||y_{i}||$ values to exactly align with the region or land boundary. The total area involved in this adjustments is very small. This step ensures that any individual generated contour looks physically realistic.

\subsection{Parametric covariance}
To allow for efficient fitting of Equation \ref{eq: piTildeNorm}, we define a parametric covariance structure. In sea ice observations, the mean and covariance of the ice-covered proportion of each line varies substantially within and across regions.  To represent these features well, we need a statistical model with a  reasonably flexible covariance structure. The values of $\tilde{\pi}_{i}$ and $\tilde{\pi}_{j}$ tend to be more similar when $L_{i}$ and $L_{j}$ are close together. So, we structure our covariance in all regions except the Central Arctic based on the differences between the indices of the lines in $\boldsymbol{L}$. 

Outside the Central Arctic, we let $\boldsymbol{\Sigma} = \boldsymbol{\Sigma}(\boldsymbol{\sigma}, \kappa)$ where $\boldsymbol{\sigma} = (\sigma_{1}, \hdots, \sigma_{n})$ and $\kappa > 0$. The element, $\boldsymbol{\Sigma}_{ij}$, in the $i$-th row and $j$-th column of this covariance is 
\begin{equation}
\label{eq: expCov}
\boldsymbol{\Sigma}_{ij} = \sigma_{i} \sigma_{j} \exp \left(- \frac{|i -j|}{\kappa}   \right). 
\end{equation}
where $| \cdot | $ denotes the absolute value. 

In the Central Arctic region, the lines are laid out in a circle so that the first and last lines are close to each other despite their indices  being far apart. Then the difference between the indices of line $i$ and line $j$ does not correspond to the distance between lines $L_{i}$ and $L_{j}$.  So we apply an alternative covariance function based on the difference between angles $\theta_{i}$ and $\theta_{j}$. Various covariance functions based on differences between angles have been proposed \citep{Gneiting2013}.   Like the other regions, we apply an exponential  covariance structure  where $\boldsymbol{\sigma} = (\sigma_{1}, \hdots, \sigma_{n})$, $\kappa > 0$, and the element in the $i$-th row and $j$-th column of the covariance matrix is
\begin{equation}
\label{eq: expCovCent}
\boldsymbol{\Sigma}_{ij} = \sigma_{i} \sigma_{j} \exp \left(- \frac{d(\theta_{i}, \theta_{j})}{\kappa}   \right) ,
\end{equation}
where $d(\theta_{i}, \theta_{j}) \in [0, \pi]$ is the smaller angle between $\theta_{i}$ and $\theta_{j}$. 

We find that an exponential covariance structure fits the data well. In particular, this covariance structure allows for the correlation to drop off rapidly as lines become farther apart, while maintaining some non-zero correlation among all lines. Allowing for the latter behavior is needed,  since some region-wide correlation would be expected given large-scale phenomena that could occur across a region, such as a particularly cold or warm month. 

\subsection{Number of lines}
Setting  $N$, the number of lines in $\boldsymbol{L}$, involves tradeoffs between accuracy and computation time. With more lines, contours can be represented in more detail. However, computation increases with the corresponding increase in the size of the covariance matrix. Insight into this tradeoff can be obtained by considering how well a set of observed contours can be approximated with only points on  lines in  $\boldsymbol{L}$. The mean difference in area between observed contours and their approximations  provides an estimate of the expected area that a generated contour built with $N$ lines cannot represent. 

In this implementation, we set $N$ based on observations of the largest region (the Central Arctic) in the forecast month with the highest variability (September).  Using data from 1995-2004, we find that approximating the observed  September contours with points on $N = 90$ lines results in area differences of approximately 2.5$\%$ of the total area while maintaining feasible computation times. We set the $N$ for other regions in proportion to how their  area compares to the area of the Central Arctic.  Using higher $N$ in general leads to slightly more accurate forecasts, since this difference in area can be modestly reduced. Moderately lowering $N$ in general will have the opposite effect.   However, for small adjustments, e.g. adding or removing 5 to 10 lines,  this pattern may not hold due to sampling error.

\subsection{Prior distribution of the mean sea ice edge}
\label{sec: muPrior}
We take a Bayesian approach to parameter estimation. We place a strong prior on $\boldsymbol{\mu}$, the mean sea ice edge, because the ensemble forecast provides considerable information about the likely location of the mean sea ice edge. 
We first reduce the systematic errors in the ensemble using Contour Shifting \citet{Director2017}, a bias-reduction method for ensemble forecasts that we now review.

\subsubsection{Contour-Shifting}
\citet{Director2017} developed Contour Shifting for the Central Arctic region, and here we extend it to all regions. We summarize the method using the notation in Section \ref{sec:  contourModel}. 

For some historical training period  preceding the forecast year, we compare the ice edge predicted by the ensemble mean forecast to the observed ice edge. For each year $j$ and line $L_{i}$ in a particular region, we record the length of the line segments extending from each point on the coastline, $B_{i}$, to the corresponding point  on the observed ice edge contour, $\tilde{S}_{i, j}^{obs}.$ We also record the lengths of the line segments from each point on the coastline, $B_{i}$, to the corresponding point on the ensemble mean ice edge, $\tilde{S}_{i, j}^{ens}.$ We denote these lengths by $||y_{i,j}^{obs}||$ and $||y_{i,j}^{ens}||$ respectively. Assuming linear change in these lengths over time, we  estimate the length to which the sea ice will extend on line $L_{i}$  at some new time point $t$ for the observed ice edge,
\begin{align}
||\hat{y}_{i, t}^{obs}|| = \hat{\alpha}_{i}^{obs} + \hat{\beta}_{i}^{obs}t ,
\end{align}
and the ensemble mean ice edge,
\begin{align}
||\hat{y}_{i, t}^{ens}|| = \hat{\alpha}_{i}^{ens} + \hat{\beta}_{i}^{ens}t.
\end{align}
Here, $\hat{\alpha}_{i}^{obs}$, $\hat{\alpha}_{i}^{ens}$, $\hat{\beta}_{i}^{obs}$, and $\hat{\beta}_{i}^{ens}$  denote fitted regression coefficients. These regressions are fit using Huber M-estimation, a form of robust linear regression \citep{Huber2011}. 

The difference between $||\hat{y}_{i, t}^{obs}||$ and $||\hat{y}_{i, t}^{ens}||$ gives the expected difference between the length predicted by the mean ensemble and the length that will be observed at time $t$. So, the forecasted length on line $L_{i}$ at time $t$ is expected to be
\begin{equation}
||\hat{y}_{i, t}^{CS}|| = ||y_{i, t}^{ens}|| + (||\hat{y}_{i, t}^{obs}|| - ||\hat{y}_{i, t}^{ens}||) ,
\end{equation}
where the superscript $CS$ indicates that a Contour Shifting adjustment has been made. 

Each adjusted length, $||y_{i,t}^{CS}||$, can be combined with the corresponding $B_{i}$ and $L_{i}$ value as in Section \ref{sec: notationAndSetup} to produce new ice edge contours. The resulting contours may have self-intersections at a small number of locations. These self-intersections can be corrected with an iterative application of the Douglas-Peuker algorithm as  described in Section \ref{sec: notationAndSetup}. Contours from ensembles that have been adjusted in this way are referred to as Contour-Shifted and have reduced systematic error compared to initial ensemble forecasts.

\subsubsection{Prior for mean proportions ice-covered}
We can now incorporate this reduced bias form of the forecasted ensemble mean  ice edge contour  into the prior for $\boldsymbol{\mu}$.  We use the following prior distribution, 
\begin{align}
\label{eq: prior mu}
\boldsymbol{\mu} \sim N(\boldsymbol{\mu}_{0}, \boldsymbol{\Lambda}_{0}),
\end{align}
where $\boldsymbol{\mu}_{0}$ is an $n \times 1$  mean vector informed by the ensemble forecast.  Let 
\begin{equation}
\pi_{i, t}^{CS} = \frac{||y_{i,t}^{CS} \cap \boldsymbol{R}_{i}||}{||\boldsymbol{R}_{i}||}. 
\end{equation}
denote the proportion of $\boldsymbol{R}_{i}$ that $y_{i,t}^{CS}$ covers  in the Contour-Shifted ensemble mean ice edge. Then for all $i$, let
\begin{equation}
\mu_{0, i} = \begin{cases}
 \text{logit} \left( ||\pi_{i, t}^{CS}|| \right) &\text { for } \epsilon \leq \pi_{i,t}^{CS} \leq 1 - \epsilon\\
 \text{logit}(\epsilon) & \text{ for } \pi_{i,t}^{CS}  < \epsilon\\
  \text{logit}(1 - \epsilon) & \text{ for } \pi_{i,t}^{CS}  > 1 -  \epsilon . \\
\end{cases}
\end{equation}

 Also, the matrix $\boldsymbol{\Lambda}_{0}$ is a $n \times n$ diagonal covariance matrix with 
\begin{align}
\boldsymbol{\Lambda}_{0, ii} =  \frac{\text{logit}(\max(\pi_{i,t}^{CS} - 0.125, \epsilon)) - \text{logit}(\min(\pi_{i,t}^{CS} + 0.125, 1 - \epsilon)))/2}{\Phi^{-1}(.995)}.
\end{align} This prior covariance treats all $\mu_{i}$ as independent. The variance for each $\mu_{i}$ is equivalent to the variance that would be obtained with 99$\%$ of the distribution's mass in the logit-transformed interval of $[\max(\pi_{i,t}^{CS} - 0.125, \epsilon), \min(\pi_{i,t}^{CS} + 0.125, 1 - \epsilon)]$. This prior variance for $\mu_{0,i}$ typically corresponds to the mean proportion being within 0.125 of the Contour-Shifted ensemble mean proportion, $\pi_{i,t}^{CS}$. The  variance is reduced if $\pi_{i,t}^{CS}$ is close to 0 or 1. (See Appendix \ref{app: sigmaInBounds} for the derivation of the standard deviation of a normal distribution that corresponds to a particular proportion of the mass of the distribution being within a set of bounds. In this case, $M = \text{logit}(\max(\pi_{i,t}^{CS} - 0.125, \epsilon))$, $m = \text{logit}(\min(\pi_{i,t}^{CS} + 0.125, 1 - \epsilon))$, and $\gamma= 0.99$.)

\subsection{Prior for Covariance}
\label{sec: SigmaPrior}
For the prior on $\boldsymbol{\Sigma}$ we only use information about the physical constraints. While ensembles have the potential to provide information about covariance, the variability of the ensembles we have analyzed do not align with the variability seen in observations.   As such, we use only physical constraints to inform our priors for the covariance parameters, $\boldsymbol{\sigma}$ and $\kappa$.

Since standard deviation values are bounded below and considerable differences in variances exist for the $\tilde{\pi}_{i}$ values, we select an independent uniform prior for each $\sigma_{i}$ such that  
\begin{align}
\sigma_{0,i} \overset{iid}{\sim} \text{Unif}(\alpha_{\sigma,0}, \beta_{\sigma,0}) ,
\end{align}

\noindent where $\alpha_{\sigma,0} = 0.01$. We bound $\sigma$ at $\alpha_{\sigma,0}$ rather than zero to avoid numerical issues when sampled $\sigma$ values approach zero. We let 
\begin{align}
\beta_{\sigma_{0, i}} =  \frac{(\text{logit}(\delta_{2} ) - \text{logit}(\delta_{1}))/2}{\Phi^{-1}(.995)}.
\end{align}

\noindent where typically $\delta_{1} = \epsilon$ and $\delta_{2} = 1 - \epsilon$. This upper bound corresponds to the standard deviation of a Gaussian distribution with 99$\%$ of the distribution's mass in the interval $(\text{logit}(\delta_{1}), \text{logit}(\delta_{2}))$.  This prior distribution ensures that the variance of the transformed proportion of ice-covered length does not substantially exceed the variance of a normal distribution that fully covers the interval of possible proportion values.  (This bound  is obtained using Appendix \ref{app: sigmaInBounds} with $M =\text{logit}(\delta_{2})$, $m = \text{logit}(\delta_{1})$, and $\gamma= 0.99$.)

 Exceptions to the typical $\delta_{1}$ and $\delta_{2}$ values are made in the Central Arctic  region where $\delta_{1} =0.15 $ and in the Greenland sea region where $\delta_{2} = 0.73$. These exceptions reflect the fact that the ice-covered proportions in these regions have never covered the full interval $[\epsilon, 1 - \epsilon]$ for any $L_{i}$ in the training observations. Even at the annual minimum, lines in the Central Arctic have never had ice coverage proportions near zero. Similarly, even at the annual maximum, lines in the Greenland sea have never had ice coverage  proportions  exceeding 0.73.  

With little information from which to anticipate how correlation decreases with distance, we use the following vague  prior for $\kappa$, 
\begin{align}
\kappa_{0}  \sim \text{Unif}(\alpha_{\kappa, 0}, \beta_{\kappa, 0}) ,
\end{align}
where $\alpha_{\kappa, 0} = 0.05$ and $\beta_{\kappa, 0} = 20$  in our implementation. This prior ensures that $\kappa$ remains positive.

\subsection{Posterior distribution}
\label{sec: postDist}
To fit this model, we need a set of observed contours drawn from the same distribution. We treat the contours in the $P$ years immediately preceding the forecast year as independent samples from the distribution of contours from which the forecast year's contour will be drawn. With this approach we are assuming that the distribution of the contours is stationary over the $P$-year period.  While this stationarity assumption is not strictly true given climate change, for decadal time scales  the effects of the climate change trend on sea ice are small relative to year-to-year variability. Therefore, we fix $P$ and assume these recent observations provide a reasonable basis on which to build a Bayesian model.  We index the years with the subscripts $j = \{1, 2, \hdots, P\}$. We denote the set of $n$ observed proportions in year $j$ by $\boldsymbol{\tilde{\pi}}_{j}$. The element $\tilde{\pi}_{ij}$ is the proportion of $\boldsymbol{R}_{i}$ that $y_{ij}$ covers in year $j$.

Combining the likelihood for the observed proportions with the prior distributions introduced in Sections \ref{sec: muPrior} and \ref{sec: SigmaPrior} gives the posterior distribution
 \begin{align} \nonumber
  \prod_{j=1}^{P} \left\{p(\boldsymbol{\tilde{\pi}}_{j}, \boldsymbol{\mu}, \boldsymbol{\sigma}, \kappa)  \right\} p(\boldsymbol{\mu}) p(\boldsymbol{\sigma}) p(\kappa) = & \prod_{j=1}^{P} \left\{ \text{N} (\boldsymbol{\tilde{\pi}}_{j}|\boldsymbol{\mu}, \boldsymbol{\Sigma} (\boldsymbol{\sigma}, \kappa)) \right\}
   \text{N}(\boldsymbol{\mu} |\boldsymbol{\mu}_{0}, \boldsymbol{\Lambda}_{0}) \times \\  \
  &\prod_{i = 1}^{n} \left\{ \text{Unif} (\sigma_{i} |\alpha_{\sigma,0}, \beta_{\sigma,0})\right\} \text{Unif}(\kappa| \alpha_{\kappa, 0}, \beta_{\kappa, 0}).
   \label{eq: posterior}
 \end{align}

\noindent The posterior means of $\boldsymbol{\mu}$ and $\boldsymbol{\Sigma}$ can be used with Equations  \ref{eq: piTildeNorm}, \ref{eq: expCov}, and \ref{eq: expCovCent} to generate $\boldsymbol{\tilde{\pi}}$.

\subsection{Model fitting}
We sample from the posterior distribution in Equation \ref{eq: posterior} for each region independently with Markov chain Monte Carlo (MCMC), using the observed sea ice in the preceding $P$ years.  MCMC diagnostics are given in Appendix Section \ref{app: mcmcDiag}. Regions that are either completely filled with sea ice or contain no sea ice in all training years are omitted from model fitting. In such cases we predict complete ice-cover or no sea ice respectively. In some months of the year, the observed proportions at the start and/or end of the fixed boundary lines are 0 or 1 for all observed $P$. We fix  these lines with proportions of 0 or 1 rather than fit them. This omission in fitting avoids estimating an excessively high $\kappa$ due to perfect correlation among the lines in these sections.  In the Central Arctic sets of lines bordering the Canadian Archipelago that have proportion 1 for all $P$ years are similarly fixed. 

We use  Metropolis steps for updating each $\mu_{i}$, $\sigma_{i}$, and $\kappa$. Normal proposals are used for each parameter at each iteration centered at their current value. For each element $i$ and iteration $t$, the log acceptance ratio for  $\mu_{i}^{(t)}$  is 
\begin{align} \nonumber
&   -\frac{1}{2}\sum_{j=1}^{P} \left(\boldsymbol{\tilde{\pi}}_{j} - \boldsymbol{\mu}^{(t)}\right)^{T}\boldsymbol{\Sigma}^{-1}(\boldsymbol{\tilde{\pi}}_{j} - \boldsymbol{\mu}^{(t)}) - \frac{1}{2}(\boldsymbol{\mu}^{(t)} - \boldsymbol{\mu}_{0})^{T}\boldsymbol{\Lambda}^{-1}(\boldsymbol{\mu}^{(t)} - \boldsymbol{\mu}_{0})\\
&  +\frac{1}{2}\sum_{j=1}^{P}(\boldsymbol{\tilde{\pi}}_{j} - \boldsymbol{\mu})^{T}\boldsymbol{\Sigma}^{-1}(\boldsymbol{\tilde{\pi}}_{j} - \boldsymbol{\mu}) + \frac{1}{2}(\boldsymbol{\mu} - \boldsymbol{\mu}_{0})^{T}\boldsymbol{\Lambda}^{-1}(\boldsymbol{\mu} - \boldsymbol{\mu}_{0}) ,
\end{align}
where $\boldsymbol{\mu}^{(t)}$ denotes the $\boldsymbol{\mu}$ vector on the $t$-th iteration with the $i$-th element proposed. For each element $i$ and iteration $t$, the log acceptance ratio for $\sigma_{i}^{(t)}$ is 
\begin{align}
\label{eq: sigmaLogAccept} \nonumber
& -\frac{n}{2} \log |(\boldsymbol{\Sigma}^{(t)})|   -\frac{1}{2}\sum_{j=1}^{P} \left(\boldsymbol{\tilde{\pi}}_{j} - \boldsymbol{\mu}\right)^{T}(\boldsymbol{\Sigma}^{(t)})^{-1}(\boldsymbol{\tilde{\pi}}_{j} - \boldsymbol{\mu}) + \mathbbm{1}[\sigma_{i}^{(t)} \in (\alpha_{\sigma,0}, \beta_{\sigma,0})]\\
& +\frac{n}{2} \log |\boldsymbol{\Sigma}|   +\frac{1}{2}\sum_{j=1}^{P} \left(\boldsymbol{\tilde{\pi}}_{j} - \boldsymbol{\mu}\right)^{T}(\boldsymbol{\Sigma}^{(t)})^{-1}(\boldsymbol{\tilde{\pi}}_{j} - \boldsymbol{\mu}) ,
\end{align}

\noindent where $\sigma_{i}^{(t)}$ denotes the proposal for the $i$-th element of $\boldsymbol{\sigma}$ on the $t$-iteration and  $\Sigma^{(t)}$ denotes the corresponding covariance matrix with $\sigma_{i}^{(t)}$  proposed.  On the $t$-th iteration, the log acceptance ratio for $\kappa^{(t)}$ is
\begin{align}
\label{eq: sigmaLogAccept} \nonumber
& -\frac{n}{2} \log |(\boldsymbol{\Sigma}^{(t)})|   -\frac{1}{2}\sum_{j=1}^{P} \left(\boldsymbol{\tilde{\pi}}_{j} - \boldsymbol{\mu}\right)^{T}(\boldsymbol{\Sigma}^{(t)})^{-1}(\boldsymbol{\tilde{\pi}}_{j} - \boldsymbol{\mu}) + \mathbbm{1}[\kappa^{(t)} \in (\alpha_{\kappa, 0}, \beta_{\kappa, 0})]\\
& +\frac{n}{2} \log |\boldsymbol{\Sigma}|   +\frac{1}{2}\sum_{j=1}^{P} \left(\boldsymbol{\tilde{\pi}}_{j} - \boldsymbol{\mu}\right)^{T}(\boldsymbol{\Sigma}^{(t)})^{-1}(\boldsymbol{\tilde{\pi}}_{j} - \boldsymbol{\mu}) ,
\end{align}
where $\kappa^{(t)}$ denotes the proposal for $\kappa$ on the $t$-th iteration and   $\boldsymbol{\Sigma}^{(t)}$ now denotes the corresponding covariance matrix with $\kappa^{(t)}$  proposed.

\section{Mixture Contour Forecasting}

\label{sec: MCF}
The contour model generally provides reasonable forecasts of the sea ice edge contour, but does have some weaknesses. 
The first is that these forecasts only focus on the contour. While the vast majority of the sea ice is contained within contiguous areas within the main sea ice edge, small areas of sea ice sometimes still form away from this main area. Areas of open water, called polynyas, also sometimes form as holes within the main sea ice area. The contour model proposed in the previous section cannot represent these features. Secondly, forecasts of this type are tied to the existing ensemble forecast, so if the initial ensemble forecast is not very accurate, such as at long lead times, the resulting forecast will not be very skillful. 
We address these weaknesses by developing a mixture model that combines the contour model with a climatological forecast that has different strengths and weaknesses. 

MCF produces a forecast distribution of ice contours that is a mixture, or weighted average, of two component distributions, the contour model introduced in the previous section and a distribution that represents recent climatology.  Here we define the climatology forecast for  each grid box as the proportion of times sea ice has been present in that grid box in the $P$ years preceding the forecast year.

The climatology forecast has different advantages and disadvantages. The climatology forecast can represent features such as holes in the sea ice or sea ice away from the main ice edge contours. However, this forecast's reliance on only the small number of observations in the past $P$ years means that it does not capture all plausible ice edge configurations. This weakness of the climatology forecast is greatest in the highly variable months around the sea ice minimum.

The weighting of the two models can be viewed as a simple case of ensemble Bayesian Model Averaging \citep{Raftery2005}. The weight is estimated by maximum likelihood using observations and predictions from preceding years.  Let $w$ be the weight of the contour model  and $1 - w$ the weight of the climatology distribution. Also, let $\gamma_{s,t}$ be  a binary indicator of whether sea ice was present in observations for some grid box  $s$ and year $t$  in the training period. Let $g_{p}(\gamma_{s,t})$ and $g_{c}(\gamma_{s,t})$ be the estimated Bernoulli probability of sea ice presence in grid box $s$ at time $t$ obtained from the contour model and the climatology respectively. In the former case, the estimated probability is the proportion of the time that grid box $s$ is within the area enclosed by the generated contours for time $t$.  The predicted probability of sea ice presence at grid box $s$ at time $t$ is then
\begin{equation}
p(\gamma_{s, t}) = w g_{p}(\gamma_{s, t}) + (1 - w)g_{c}(\gamma_{s, t})
\end{equation}

 Assuming that errors in space and time are independent, the corresponding log-likelihood is
\begin{align}
l(w) = \sum_{t} \sum_{s} \log \{ w \, a_{s} g_{p}(\gamma_{s,t}) + (1 - w)a_{s}g_{c}(\gamma_{s,t})\}.
\end{align}
The variable $a_{s}$ is the proportion of the entire area that is in grid box $s$, i.e, $\sum_{s} a_{s} = 1$. The use of $a_{s}$ accounts for the fact that the grid boxes do not all have the same area.   Assuming spatial and temporal independence is almost certainly inaccurate; however, \citet{Raftery2005} found in a similar case that results were not particularly sensitive to this assumption.

To maximize this log-likelihood we use the Expectation-Maximization algorithm \citep{Dempster1977}.  This optimization algorithm can be applied in situations where if some unobserved quantity were known, estimation of the variable(s) of interest would be simple. In this case estimating $w$ would be straightforward if we knew for every grid box and time point whether the climatology model or the contour model estimated the observed sea ice presence more accurately. So, we introduce the latent variable $z_{p,s,t},$ which has value 1 if the contour model is the best forecast for grid box $s$ in year $t$ and 0 otherwise.  The variable $z_{c,s,t}$ is defined analogously for climatology. Note that only one of the parameters $z_{c,s,t}$ or $z_{p,s,t}$ could truly be 1; but for estimation these parameters can take any value in the interval $[0, 1]$.  Also, note $\hat{z}_{p,s,t} = 1  - \hat{z}_{c,s,t}$. Then the E-step is 
\begin{align}
\hat{z}_{p,s,t}^{(j)} = \frac{w^{(j-1)}a_{s}g_{p}(\gamma_{s,t})}{w^{(j-1)}a_{s}g_{p}(\gamma_{s,t}) + (1-w^{(j-1)})a_{s}g_{c}(\gamma_{s,t})} ,
\end{align}
and the M-step is
\begin{align}
\label{eq: M-step}
w^{(j)} = \frac{\sum_{t} \sum_{s}a_{s} \hat{z}^{(j)}_{p, s,t}}{\sum_{t}\sum_{s}a_{s}} 
\end{align}
for the $j$-th iteration.  To avoid degeneracies, any $(s, t)$ pairs where $g_{p}(\gamma_{s,t}) = g_{c}(\gamma_{s,t})$ are omitted from this maximization. Therefore, the denominator in Equation \ref{eq: M-step} may be unequal to the number of years in the training period.

\section{Method evaluation}
\label{sec: methodEvaluation}
\subsection{Model outputs and observations}
All post-processing methods are applied to the fifth generation of the  European Centre for Medium-Range Weather Forecasts (ECMWF) seasonal forecasting system (SEAS5) \citep{Johnson2019, ECMWF2017}. The relevant sea ice concentration model output can be downloaded from the Copernicus Climate Change Service Climate Data Store  \citep{C3S2019}. Among a set of publicly available ensembles without post-processing, ECMWF has been shown to be generally the most skillful \citep{Zampieri2018}. The 25-member ensemble ECMWF forecasts are initialized monthly and extend to 215 days.   Model output was regridded to the National Snow and Ice Data Center Polar stereographic grid with an approximately 25km by 25km grid \citep{nsidc2016} using a nearest-neighbors method \citep{Zhuang2018}.  Daily model output was averaged to monthly to match observations. 

We evaluate forecast accuracy by comparing predictions to a monthly-averaged sea ice concentration produced from the National Aeronautics and Space Administration satellites Nimbus-7 SMMR and DMSP SSM/I-SSMIS. This data can be downloaded from the National Snow and Ice Data Center \citep{Comiso2000}. Grid boxes with sea ice concentrations of at least 15$\%$ are treated as having sea ice present. Otherwise grid boxes are treated as not containing sea ice.
 
We evaluate forecasting skill for monthly-averaged sea ice at lead times of 0.5 months to 6.5 months in the year 2008-2016.  We report lead times treating the monthly mean as the halfway point within a month. For example, the 0.5-month lead forecast for January refers to the average of the first 31 days of a forecast initialized on January 1st. Grid boxes that are coded as land in the observations, the ensemble, or the \textit{IceCast} R package \citep{Director2020} are treated as land.

The forecasts previously described are summarized in lines 1-4 of Table \ref{tab: forecastTypes}.  Beginning in 1993, all years preceding the forecast year are used in fitting Contour Shifting. A ten-year rolling window is used to fit the statistical model for generating contours and in the climatology forecast weighted in MCF. The time length of ten years is used, since recent analyses have shown that climatology computed over a ten-year period provides reasonably accurate sea ice forecasts. Such forecasts become nearly as accurate as ECMWF  ensemble forecasts at lead times of 1.5 months \citep{Zampieri2018}. Slight changes in the number of years used for this purpose is unlikely to affect the results. However, using many fewer years would not provide enough samples to fit the parameters accurately. Using a much bigger number of years would also degrade performance. Because of the rapid reduction in Arctic sea ice area, contours  from past decades will differ notably from recent ice edge contours. A three-year rolling window is used to determine the weights in MCF. Performance accuracy is generally insensitive to this choice (see Appendix \ref{sec: trainLengths}). One hundred contours are generated for each forecast.

\subsection{Reference forecasts}

\begin{table}[thb]
\caption{Summary of forecast types evaluated.  Probabilistic forecasts give estimates in the interval $[0, 1]$ and binary forecasts indicate predicted sea ice presence}
\centering\renewcommand\cellalign{lc}
\setcellgapes{3pt}\makegapedcells
\begin{tabular}{|l|c|l|} \hline
\label{tab: forecastTypes}
\textbf{Forecast} & \textbf{Probabilistic} & \textbf{Binary} \\ \hline
\makecell{\textbf{Ensemble}} &\makecell{proportion of ensemble members \\ predicting sea ice} & \makecell{indicator of whether median  \\ensemble member predicts sea ice}   \\ \hline
\makecell{\textbf{Contour}} & \makecell{ensemble mean forecast  \\bias-corrected with \\ Contour-Shifting  and calibrated by \\generating contours } & \makecell{ensemble mean forecast\\ adjusted with Contour-Shifting}\\
\hline
\textbf{Climatology} & \makecell{proportion of observations in the \\10  years\\ preceding the forecast year\\ that contain sea ice} & \makecell{indicator of whether at least five\\ of the ten  years preceding\\ the forecast year contained sea ice }\\
\hline
\makecell{\textbf{Mixture Contour} \\ \textbf{Forecast (MCF)}} & \makecell{forecast  formed by weighting\\ probability densities  from\\ climatology and the contour model} & \makecell{ indicator of whether forecast \\formed by weighting probability \\densities from climatology \\ and the contour model predicts \\sea ice with $p \geq 0.5$} \\
\hline
\makecell{\textbf{Trend Adjusted} \\ \textbf{Quantile Mapping} \\\textbf{(TAQM)}}& \makecell{ensemble post-processed using \\technique in \citet{Dirkson2019}}  & NA\\
\hline
\textbf{Damped Persistence} & NA & \makecell{indicator of whether predicted sea ice\\ concentration from a damped \\persistence  forecast is at least 0.15 \\(modified  from \citet{Wayand2019})}\\
\hline
\end{tabular}
\end{table}

We compare our results to two additional reference forecasts summarized in lines 5-6 of Table \ref{tab: forecastTypes}.  Trend Adjusted Quantile Mapping (TAQM)  is another recently developed statistical post-processing method for sea ice \citep{Dirkson2019}. TAQM fits a parametric probability distribution to ensemble model output and applies a specialized version of quantile mapping  to produce probabilistic forecasts of sea ice concentration. TAQM does not predict the probability of sea ice presence directly, but  \citet{Dirkson2019} do use the resulting distribution of the sea ice concentration to predict the probability of  sea ice presence (concentration of at least 15$\%$).   

We also compute a damped persistence forecast in a manner similar to \citet{Wayand2019}. Damped persistence forecasts estimate the sea ice concentration in forecast month $m$ using linear regression and the observed sea ice concentration in the initialization month $i$.  For each grid box, the concentration for month $m$ in year $t$, denoted by $C_{m,t}$, is estimated as
\begin{align}
\label{eq: dPersis}
\hat{C}_{m, t} = \hat{\beta}_{m} t + (C_{i, t_{i}} - \hat{\beta}_{i}t_{i}) \hat{\rho},
\end{align}
where $C_{i, t_{i}}$ is the observed concentration in the $i$-th initialization month, $\hat{\beta}_{m}$ and $\hat{\beta}_{i}$ are the coefficients for linear regressions of $C_{m}$ and $C_{i}$ on year, and $\hat{\rho}$ is the empirical correlation of $C_{m}$ and $C_{i}$  estimated from past observations. If all values of $C_{m}$ and/or $C_{i}$ in the training period are the same, the empirical correlation is undefined. In these cases,  we set $\hat{\rho}$ to 0, which makes Equation \ref{eq: dPersis} equivalent to linear regression. When $i \leq m$,  $t_{i} = t$,  otherwise $t_{i} = t - 1$.  Grid boxes with predicted concentration of at least 0.15 are forecasted to contain sea ice. Observations beginning in 1981 and extending up to the initialization time are used in fitting.

\subsection{Visualizing forecasts}

 \begin{figure}[!htbp]
		\centering
		\includegraphics[height = .8\textheight]{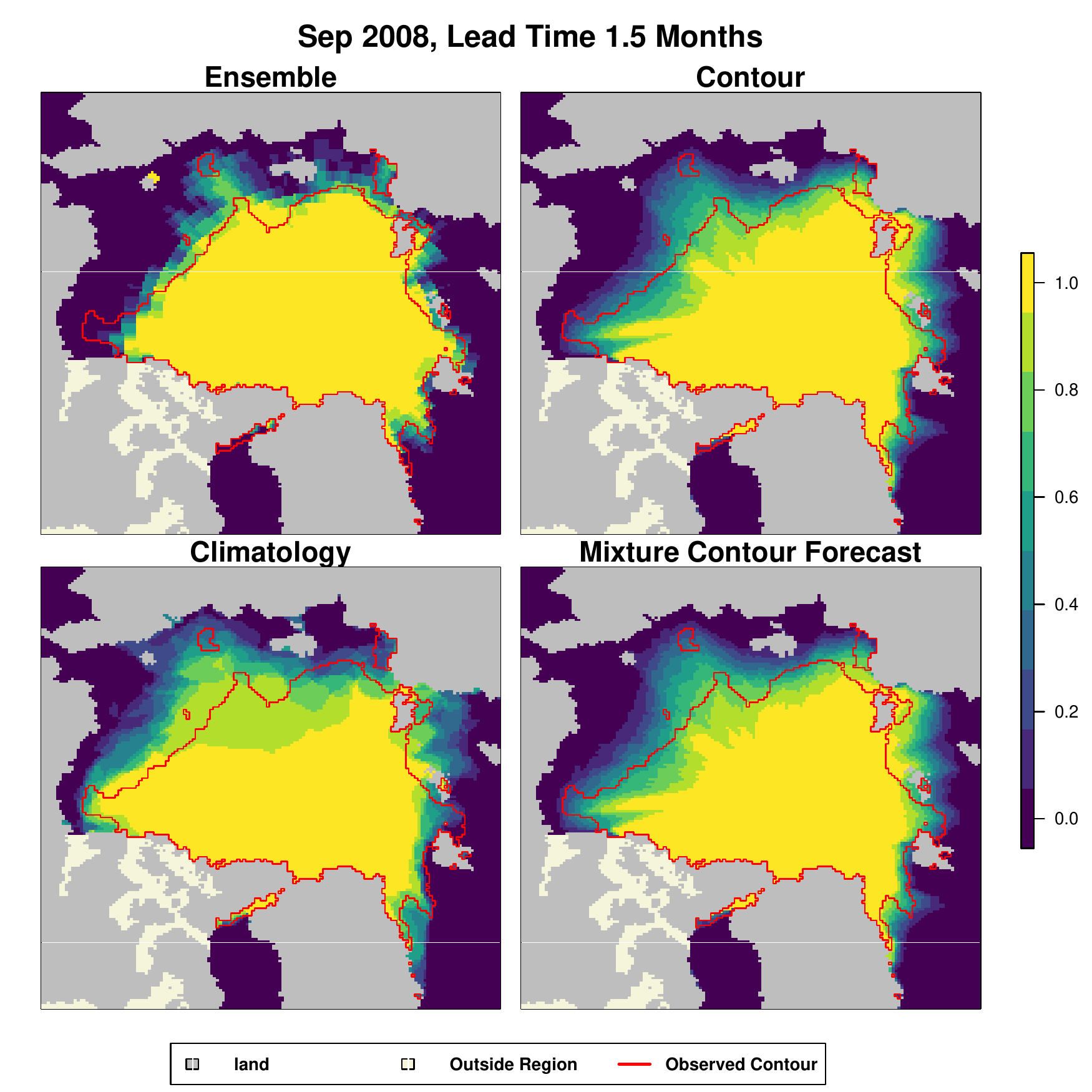} 
		\caption{Forecasts of the probability of sea ice presence for September 2008 using different methods. The forecasts are described in Table \ref{tab: forecastTypes}. The red line is the observed sea ice edge contour and grey areas are land. For MCF, the observed ice edge is almost completely within areas with positive probability and has little area where sea ice is was predicted with probability 1 but sea ice was not present in observations. In contrast, the observed sea ice edge more often goes through regions with zero probability in the ensemble and climatology forecasts. In the climatology forecast, there is also the most area where sea ice is predicted with probability 1 but is not present in observations. }
		\label{fig: visSep}
\end{figure}

Uncertainty information is needed for maritime planning to adequately evaluate risks and benefits. Like \citet{Gneiting2007}, we consider accurate model calibration to be vital for probabilistic forecasts. We illustrate the importance of calibration in this context with Figure \ref{fig: visSep}, which shows samples of four  probabilistic  forecasts for September 2008. The corresponding observed sea ice edge for September 2008 is also plotted for reference. Figure \ref{fig: visSep} illustrates the types of forecasting errors that can occur when forecasts are not calibrated. Specifically, events with low predicted probability occur more often than expected, and/or events with high predicted probability occur less often than expected.  

For the contour model and MCF, the observed contour is almost entirely contained within regions with positive probability and has only small areas where sea ice is predicted with probability 1 but sea ice is not present in observations. The MCF forecast is slightly more variable than the contour model, reflecting its weighting with climatology. Since $w$ is high for September at a 1.5-month lead time, the difference between the contour model and MCF is small. In cases where $w$, the weight on the contour model, is lower, the difference between the contour model and MCF may be more substantial. For the climatology and ensemble forecasts,  the observed contour goes through some regions with zero probability, suggesting that these forecasts are not sufficiently variable. For these forecasts, there are areas where sea ice is predicted with probability 0, but sea ice is observed. Discrepancies like these between the forecasted probability and what will likely occur makes maritime planning and risk mitigation with these latter types of forecasts difficult.

\subsection{Assessing Calibration}

We now evaluate model calibration for the probabilistic forecasts. We evaluate calibration with reliability diagrams that plot the forecasted probability of observing sea ice against the proportion of times sea ice was observed.  A perfectly calibrated forecast would have all points on the $y = x$ line, i.e. grid boxes forecasted to contain sea ice with a given probability actually contain sea ice the same proportion of the time. So, the closer the points lie to the $y = x$ line, the better calibrated the forecast is.

\begin{figure}[!htb]
		\centering
		\includegraphics[width= .95\textwidth]{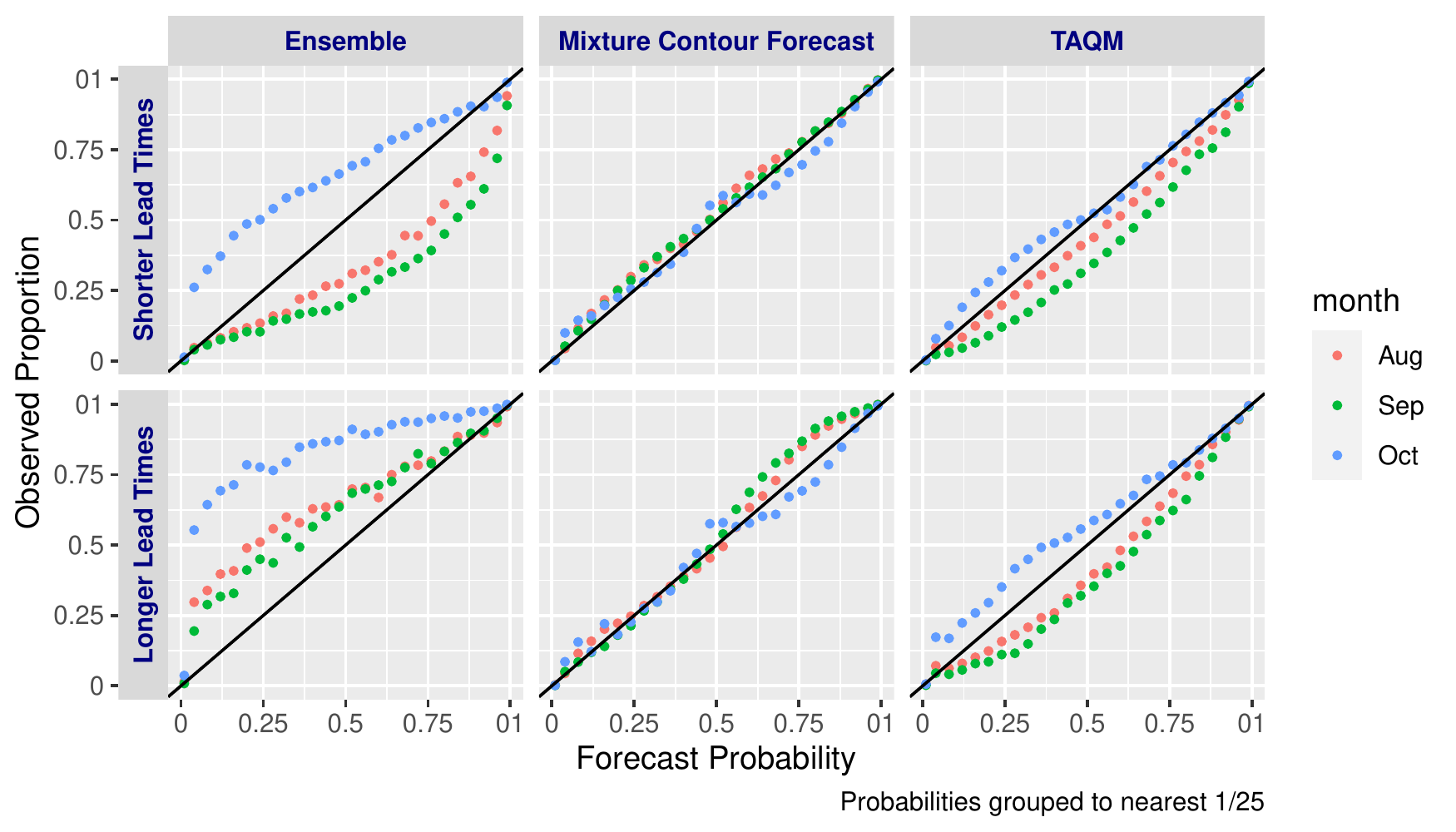} 
		\caption{The average proportion of the time sea ice was present plotted against the predicted probability of sea ice presence for the raw ECMWF forecasts (left), after post-processing with MCF (middle), and TAQM (right). Forecasts are grouped into lead times of 0.5 and 1.5 months (top) and 2.5-6.5 months (bottom). A perfectly calibrated forecast would have all points on the diagonal $y = x$ line. }
			\label{fig: calib_ASO}
\end{figure}

 Shipping varies seasonally in the Arctic, with more shipping in months around the annual sea ice minimum in September \citep{Ellis2009}, so we  emphasize performance in these peak shipping months. In Figure \ref{fig: calib_ASO} we show the reliability diagrams for the peak-shipping months  for the probabilistic forecasts. Predictions from MCF are substantially better calibrated than the ensemble and better calibrated than TAQM during these months, especially at short lead times. Figures \ref{fig: calibL2} and \ref{fig: calibG2} in Appendix \ref{app: addFigs} show that MCF always improves calibration over the unadjusted ensemble and generally improves calibration compared to TAQM.  We note that MCF has been designed specifically for predicting sea ice presence, while TAQM addresses the more general goal of forecasting sea ice concentration. The difference in calibration performance between MCF and TAQM highlights the benefit for maritime planning of having a method focused exclusively on predicting sea ice presence. TAQM remains valuable for its broader applicability.

\subsection{Assessing accuracy}

\begin{figure}[htbp!]
\begin{subfigure}{1.2\textwidth}
	\centering
	\includegraphics[width= .85\textwidth]{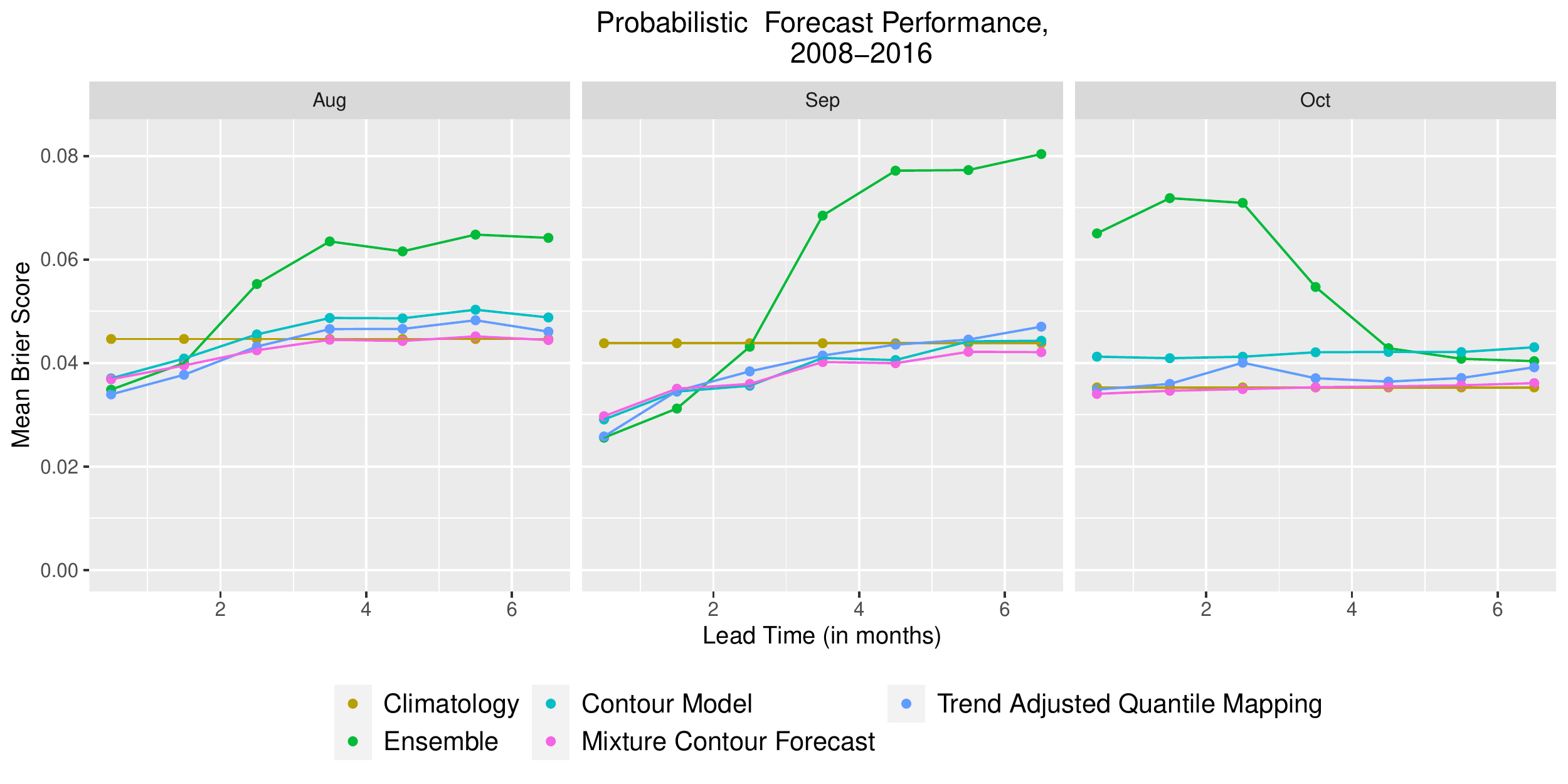} 
\end{subfigure}
\begin{subfigure}{1.2\textwidth}
	\centering
	\includegraphics[width= .85\textwidth]{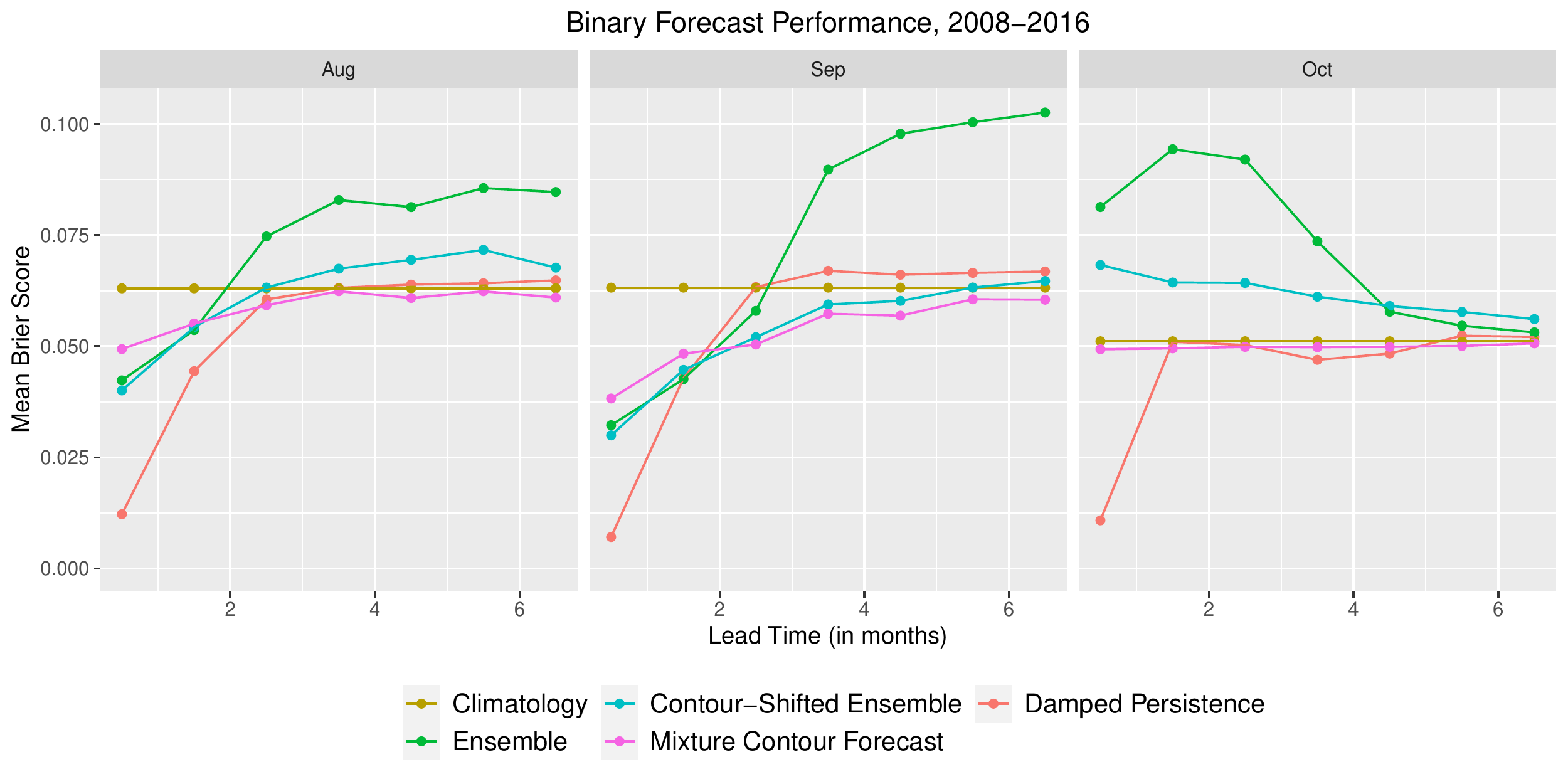} 
\end{subfigure}
		\caption{Top: Average Brier scores by month for the test years 2008-2016 for the probabilistic forecasts. The Brier Score for each grid box is weighted based on its area.  Forecasts are described in Table \ref{tab: forecastTypes}. Bottom: As above, but for the binary forecasts.}
		\label{fig: brier}

\end{figure}

We evaluate forecast accuracy using Brier scores \citep{Brier1950}. We compute average area-weighted Brier scores over the $T = 9$ years in the test set as
\begin{equation}
\frac{\sum_{t} \sum_{s}a_{s}(f_{s,t} - o_{s,t})^{2}}{T},
\end{equation}
where $f_{s,t}$ and $o_{s,t}$ denote the forecast and observation in grid box $s$ in year $t$ respectively. The value $a_{s}$ is the proportion of the total area that is in grid box $s$. The observed value is 1 when the sea ice concentration is at least 0.15, and 0 otherwise. For probabilistic forecasts, $f_{i,j} \in [0, 1]$ and  for binary forecasts, $f_{i,j} \in \{0, 1\}$. 

In Figure \ref{fig: brier}, we plot the average Brier score in peak-shipping months by lead time for the probabilistic forecasts. The ensemble forecasts typically have increasing Brier scores as lead time increases. Our contour model generally improves forecast accuracy and MCF improves accuracy further.  As lead time increases, MCF's performance converges to equal or better performance than climatology. TAQM also generally improves accuracy of forecasts. 

Figures  \ref{fig: brier_prob_overall} and \ref{fig: brier_prob_seas} in Appendix \ref{app: addFigs} show that TAQM and MCF have similar overall accuracy, but that the pattern of their performance by lead time and month varies. For peak shipping months, MCF outperforms TAQM, suggesting that our specialized modeling of the sea ice edge has benefits for maritime planning use. For other applications, more general techniques like TAQM may be more appropriate.  For the shortest lead time of 0.5 months, the damped persistence forecast performs best, but its skill decays rapidly with lead time. The performance of the damped persistence forecast indicates that there could be a role for the current observed state of the sea ice in forecasting, but that the role would need to be restricted to very short lead times. In summary, MCF provides the best calibrated forecasts year round and accurate forecasts during peak-shipping months.

\subsection{Binary Forecasts}
We also briefly assess binary forecasts with the bottom panel in  Figure \ref{fig: brier}. Binary forecasts are inherently poorly calibrated, and so are not optimal, but can be useful in method assessment. The Contour-Shifted forecast clearly improves accuracy compared to the ensemble. In other words, Contour Shifting does reduce some systematic bias that affects typical ensembles. Binary MCF performs similarly to the Contour-Shifted ensemble in general, but MCF substantially outperforms the Contour-Shifted ensemble when the ensemble forecast is poor. This case illustrates that the adaptive weighting provided by MCF is valuable when issuing binary forecasts as well as probabilistic forecasts. Brier scores for binary forecasts for all seasons are in \ref{fig: brier_bin_seas} in Appendix \ref{app: addFigs}.

\subsection{Understanding mixture weights}

\begin{figure}[!htbp]
		\centering
		\includegraphics[width= .75\textwidth]{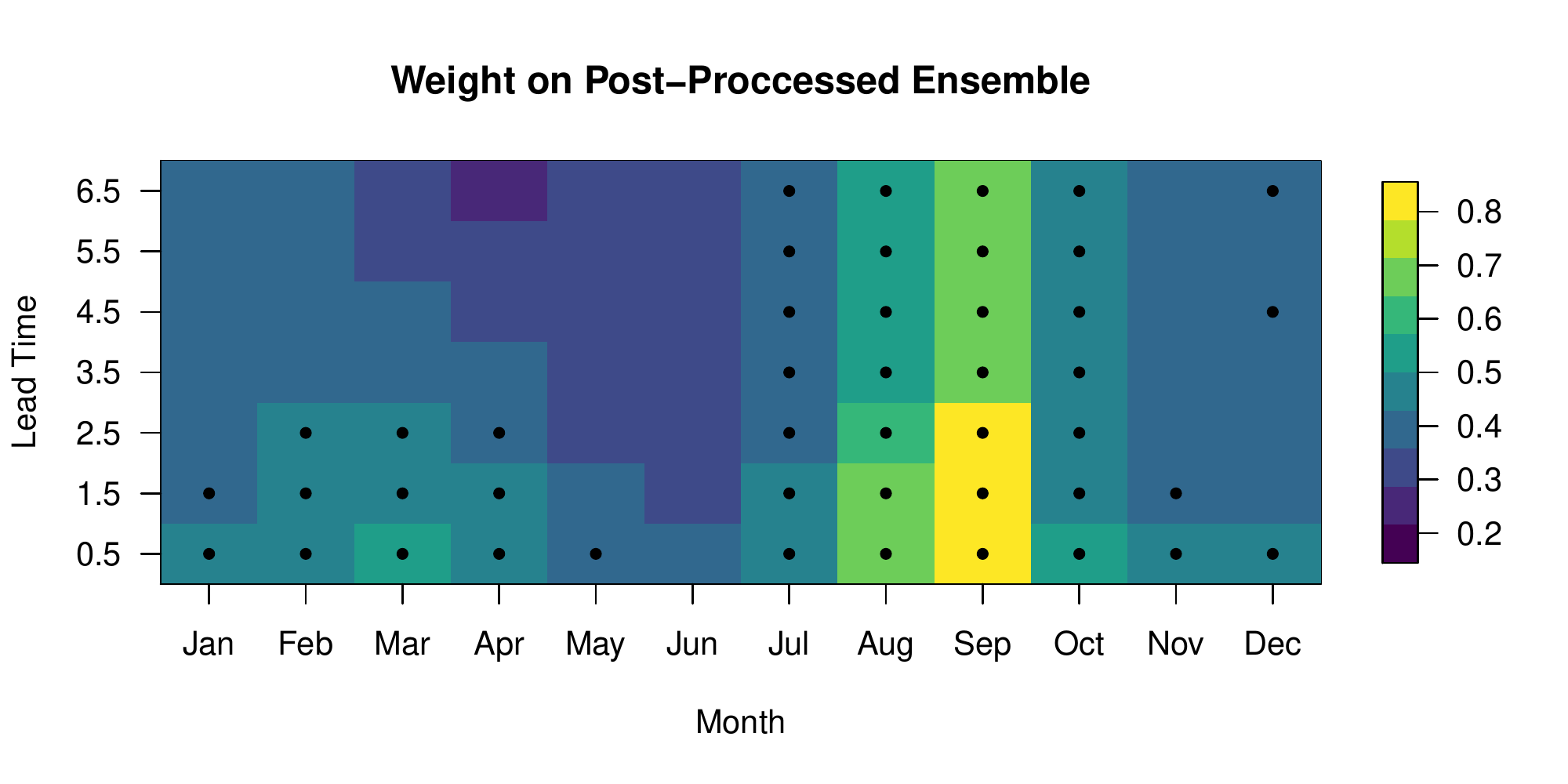} 
		\caption{Weight on the contour model by month and lead time. A black dot indicates that the weight is at least 0.4. The contour gets more weight at short lead times and in months near the sea ice minimum. The weight on climatology is equal to one minus the weight on the contour model.}
		\label{fig: avWeights}
\end{figure}

Since MCF is fitted separately for each forecast month and lead time, we can examine how the weights on the contour model and climatology forecast vary between months and lead times. Figure \ref{fig: avWeights} shows the average weight placed on the contour model for the years in the test set. High weights typically occur at short lead times, reflecting the fact that the ensemble typically has the most skill soon after it is initialized. High weights also occur in months around the sea ice minimum in September. These are periods of high year-to-year variability, so climatology tends to perform poorly and the ensemble's ability to simulate evolving physical conditions becomes more important.

\section{Discussion}
\label{sec: discuss}

We have introduced the Mixture Contour Forecasting method for issuing probabilistic sea ice forecasts. MCF forecasts are probabilistic and well calibrated, meaning that their predicted probability of sea ice presence at a given location approximately matches the proportion of times sea ice will be observed at these locations. At most lead times and forecast months, probabilistic MCF forecasts are also as or more accurate than the raw ECMWF ensemble and the other post-processed and statistical forecasts.  

Because MCF provides well calibrated and relatively accurate forecasts,  MCF's use has the potential to increase operational sea ice forecasting skill and thereby improve maritime planning in the Arctic. As Arctic routes are planned, the risk of encountering sea ice where it is not expected must be weighed against the cost savings of a shorter route.  Vessels in the Arctic have an ice classification that says where they can legally and safely travel.  For vessels that are easily damaged in sea ice, encountering any sea ice poses great risk. In contrast, ships that are designed to travel safely through sea ice may gain speed and efficiency by avoiding sea ice, but do not face danger if they encounter it. 

Our model evaluation weights both types of misclassification errors equally (predicting the presence of sea ice when it was not observed and predicting the absence of sea ice when it was observed.) However, the probabilistic forecasts provided by MCF do allow us to account for different costs of the two types of error. In particular, MCF opens up the possibility of planning routes using  decisions rules that incorporate the probability of sea ice presence. For example, a ship that has high risk of damage when traversing sea ice might elect to only consider routes through areas with very low probability of sea ice.

We have also developed a framework for directly modeling contours. While forecasts could likely be made with field-based geostatistical models \citep[e.g.,][]{Zimmerman2010} or by identifying the exceedance level contours estimated from fields  \citep{Bolin2015, French2016}, these approaches may have limitations for this application. Most of the error in sea ice forecasts occurs in the region where a rapid transition from fully ice-covered regions to open water occurs \citep{Tietsche2014}. Whether sea ice will be found in grid boxes in the interior of the sea ice region and far from the sea ice edge is essentially known in advance. So placing the majority of the computational cost and modeling effort on the boundary is advantageous. MCF provides a framework for modeling that could be extended to other situations where the boundary is of interest.

 As implemented in this paper,  estimates of the covariance of the sea ice edge are based on the covariance estimated from the preceding years. These estimates are therefore independent of the covariance of the sea ice edge in the ensemble members.  However, the ensemble could plausibly give information about the expected covariance that could not be obtained from past observations. For example, sea ice is expected to continue to become thinner. Thinner sea ice is more affected by variation in meteorological conditions, so the variance of sea ice extent will likely increase \citep{Holland2011}. Effects like these are captured by the ensemble, but are not in past observations. As such, incorporating the covariance in the ensemble could further improve forecast skill. However, the spread of the ensemble does align with observed variability and the relationship between variability in observations and variability in the ensemble is inconsistent both spatially and temporally. Thus assessment of when the ensemble covariance is informative and how it relates to the observed covariance is needed before it will be feasible to incorporate the ensemble covariance into MCF. 
     
The ECMWF ensemble used in Section \ref{sec: methodEvaluation} is not the only ensemble prediction system. The post-processing techniques developed in this paper could be directly applied to other ensembles, since they do not use any specific features of the ECMWF ensemble. However, model biases and calibration issues vary, so exact performance  would need to be assessed. Different ensembles also vary by which forecast months they perform well in, and vary more by the extent to which skill declines with lead time \citep{Zampieri2018}. Thus extending MCF to use multiple ensemble members as has been done for other meteorological variables could provide further skill  \citep[e.g.,][]{Raftery2005, Dirkson2019b}.

Our analysis of sea ice forecasting highlights situations where statistical post-processing can provide value. Many aspects of physical processes are known to evolve following well established equations. Such information can only be crudely approximated with a purely observational data-driven approach. On the other hand,  physical models are often biased or poorly calibrated, and statistical post-processing methods can be effective in remedying these problems. Combining the strengths of physical  and statistical modeling can create predictions that are more accurate than either modeling framework alone.


\clearpage
\bibliographystyle{apalike}
\bibliography{references}

\begin{thebibliography}{}

\bibitem[Blanchard-Wrigglesworth et~al., 2015]{Blanchard2015}
Blanchard-Wrigglesworth, E., Cullather, R.~I., Wang, W., Zhang, J., and Bitz,
  C.~M. (2015).
\newblock Model forecast skill and sensitivity to initial conditions in the
  seasonal {Sea Ice Outlook}.
\newblock {\em Geophysical Research Letters}, 42(19):8042--8048.

\bibitem[Bolin and Lindgren, 2015]{Bolin2015}
Bolin, D. and Lindgren, F. (2015).
\newblock Excursion and contour uncertainty regions for latent gaussian models.
\newblock {\em Journal of the Royal Statistical Society: Series B (Statistical
  Methodology)}, 77(1):85--106.

\bibitem[Brier, 1950]{Brier1950}
Brier, G. (1950).
\newblock Verification of forecasts expressed in terms of probability.
\newblock {\em Monthly Weather Review}, 78(1):1--3.

\bibitem[Bushuk et~al., 2017]{Bushuk2017}
Bushuk, M., Msadek, R., Winton, M., Vecchi, G.~A., Gudgel, R., Rosati, A., and
  Yang, X. (2017).
\newblock Skillful regional prediction of {Arctic} sea ice on seasonal
  timescales.
\newblock {\em Geophysical Research Letters}, 44(10):4953--4964.

\bibitem[Cavalieri and Parkinson, 2012]{Cavalieri2012}
Cavalieri, D.~J. and Parkinson, C.~L. (2012).
\newblock {Arctic} sea ice variability and trends, 1979-2010.
\newblock {\em The Cryosphere}, 6(4):881.

\bibitem[Chevallier et~al., 2013]{Chevallier2013}
Chevallier, M., Salas~y M{\'e}lia, D., Voldoire, A., D{\'e}qu{\'e}, M., and
  Garric, G. (2013).
\newblock Seasonal forecasts of the {pan-Arctic} sea ice extent using a
  {GCM-based} seasonal prediction system.
\newblock {\em Journal of Climate}, 26(16):6092--6104.

\bibitem[Comiso, 2017]{Comiso2000}
Comiso, J. (2017).
\newblock Bootstrap sea ice concentrations from {Nimbus-7 SMMR} and {DMSP
  SSM/I-SSMIS}. version 3.

\bibitem[Comiso et~al., 2008]{Comiso2008}
Comiso, J.~C., Parkinson, C.~L., Gersten, R., and Stock, L. (2008).
\newblock Accelerated decline in the {Arctic} sea ice cover.
\newblock {\em Geophysical Research Letters}, 35(1).

\bibitem[{Copernicus Climate Change Service}, 2019]{C3S2019}
{Copernicus Climate Change Service} (2019).
\newblock {C}opernicus climate change service climate data store.
\newblock \url{https://cds.climate.copernicus.eu}.

\bibitem[Dempster et~al., 1977]{Dempster1977}
Dempster, A.~P., Laird, N.~M., and Rubin, D.~B. (1977).
\newblock Maximum likelihood from incomplete data via the {EM} algorithm.
\newblock {\em Journal of the Royal Statistical Society: Series B
  (Methodological)}, 39(1):1--22.

\bibitem[Director et~al., 2017]{Director2017}
Director, H.~M., Raftery, A.~E., and Bitz, C.~M. (2017).
\newblock Improved sea ice forecasting through spatiotemporal bias correction.
\newblock {\em Journal of Climate}, 30(23):9493--9510.

\bibitem[Director et~al., 2020]{Director2020}
Director, H.~M., Raftery, A.~E., and Bitz, C.~M. (2020).
\newblock Icecast: Apply statistical post-processing to improve sea ice
  predictions.
\newblock R package version 3.0.0,
  \url{https://github.com/hdirector/IceCastV3}.

\bibitem[Dirkson et~al., 2019a]{Dirkson2019b}
Dirkson, A., Denis, B., and Merryfield, W.~J. (2019a).
\newblock A multimodel approach for improving seasonal probabilistic forecasts
  of regional arctic sea ice.
\newblock {\em Geophysical Research Letters}.

\bibitem[Dirkson et~al., 2019b]{Dirkson2019}
Dirkson, A., Merryfield, W.~J., and Monahan, A.~H. (2019b).
\newblock Calibrated probabilistic forecasts of {Arctic} sea ice concentration.
\newblock {\em Journal of Climate}, 32(4):1251--1271.

\bibitem[Douglas and Peucker, 1973]{Douglas1973}
Douglas, D.~H. and Peucker, T.~K. (1973).
\newblock Algorithms for the reduction of the number of points required to
  represent a digitized line or its caricature.
\newblock {\em Cartographica: the international journal for geographic
  information and geovisualization}, 10(2):112--122.

\bibitem[Ellis and Brigham, 2009]{Ellis2009}
Ellis, B. and Brigham, L. (2009).
\newblock {Arctic} marine shipping assessment 2009 report.

\bibitem[{European Centre for Medium-Range Weather Forecasts}, 2017]{ECMWF2017}
{European Centre for Medium-Range Weather Forecasts} (2017).
\newblock {ECMWF} {SEAS5} user guide.
\newblock
  \url{https://www.ecmwf.int/sites/default/files/medialibrary/2017-10/System5_guide.pdf}.

\bibitem[French and Hoeting, 2016]{French2016}
French, J.~P. and Hoeting, J.~A. (2016).
\newblock Credible regions for exceedance sets of geostatistical data.
\newblock {\em Environmetrics}, 27(1):4--14.

\bibitem[Gneiting, 2013]{Gneiting2013}
Gneiting, T. (2013).
\newblock Strictly and non-strictly positive definite functions on spheres.
\newblock {\em Bernoulli}, 19(4):1327--1349.

\bibitem[Gneiting et~al., 2007]{Gneiting2007}
Gneiting, T., Balabdaoui, F., and Raftery, A.~E. (2007).
\newblock Probabilistic forecasts, calibration and sharpness.
\newblock {\em Journal of the Royal Statistical Society: Series B (Statistical
  Methodology)}, 69(2):243--268.

\bibitem[Guemas et~al., 2016]{Guemas2016}
Guemas, V., Blanchard-Wrigglesworth, E., Chevallier, M., Day, J.~J.,
  D{\'e}qu{\'e}, M., Doblas-Reyes, F.~J., Fu{\v{c}}kar, N.~S., Germe, A.,
  Hawkins, E., Keeley, S., et~al. (2016).
\newblock A review on {Arctic} sea-ice predictability and prediction on
  seasonal to decadal time-scales.
\newblock {\em Quarterly Journal of the Royal Meteorological Society},
  142(695):546--561.

\bibitem[Holland et~al., 2011]{Holland2011}
Holland, M.~M., Bailey, D.~A., and Vavrus, S. (2011).
\newblock Inherent sea ice predictability in the rapidly changing {Arctic}
  environment of the {Community Climate System Model}, version 3.
\newblock {\em Climate Dynamics}, 36(7-8):1239--1253.

\bibitem[Huber, 2011]{Huber2011}
Huber, P.~J. (2011).
\newblock {\em Robust Statistics}.
\newblock Springer.

\bibitem[Johnson et~al., 2019]{Johnson2019}
Johnson, S.~J., Stockdale, T.~N., Ferranti, L., Balmaseda, M.~A., Molteni,
  F.and~Magnusson, L., Tietsche, S., Decremer, D., Weisheimer, A., Balsamo, G.,
  Keeley, S. P.~E., Mogensen, K., Zuo, H., and {Monge-Sanz}, B.~M. (2019).
\newblock {SEAS5}: the new {ECMWF} seasonal forecast system.
\newblock {\em Geoscientific Model Development}, 12(3).

\bibitem[Melia et~al., 2016]{Melia2016}
Melia, N., Haines, K., and Hawkins, E. (2016).
\newblock Sea ice decline and 21st century {trans-Arctic} shipping routes.
\newblock {\em Geophysical Research Letters}, 43(18):9720--9728.

\bibitem[Msadek et~al., 2014]{Msadek2014}
Msadek, R., Vecchi, G.~A., Winton, M., and Gudgel, R.~G. (2014).
\newblock Importance of initial conditions in seasonal predictions of {Arctic}
  sea ice extent.
\newblock {\em Geophysical Research Letters}, 41(14):5208--5215.

\bibitem[{National Snow and Ice Data Center}, 2017]{nsidc2016}
{National Snow and Ice Data Center} (2017).
\newblock Region mask for the {Northern} {Hemisphere}.
\newblock \url{http://nsidc.org/data/polar-stereo/tools_masks.html}.

\bibitem[Plummer et~al., 2006]{Plummer2006}
Plummer, M., Best, N., Cowles, K., and Vines, K. (2006).
\newblock Coda: Convergence diagnosis and output analysis for mcmc.
\newblock {\em R News}, 6(1):7--11.

\bibitem[Raftery et~al., 2005]{Raftery2005}
Raftery, A.~E., Gneiting, T., Balabdaoui, F., and Polakowski, M. (2005).
\newblock Using {Bayesian} model averaging to calibrate forecast ensembles.
\newblock {\em Monthly Weather Review}, 133(5):1155--1174.

\bibitem[Raftery and Lewis, 1992]{Raftery1992}
Raftery, A.~E. and Lewis, S.~M. (1992).
\newblock Practical {Markov} chain {Monte Carlo}: comment: one long run with
  diagnostics: implementation strategies for {Markov} chain {Monte Carlo}.
\newblock {\em Statistical Science}, 7(4):493--497.

\bibitem[Raftery and Lewis, 1995]{Raftery1995}
Raftery, A.~E. and Lewis, S.~M. (1995).
\newblock The number of iterations, convergence diagnostics and generic
  metropolis algorithms.
\newblock {\em Practical {Markov} Chain {Monte Carlo}}, 7(98):763--773.

\bibitem[Sigmond et~al., 2013]{Sigmond2013}
Sigmond, M., Fyfe, J.~C., Flato, G.~M., Kharin, V.~V., and Merryfield, W.~J.
  (2013).
\newblock Seasonal forecast skill of {Arctic} sea ice area in a dynamical
  forecast system.
\newblock {\em Geophysical Research Letters}, 40(3):529--534.

\bibitem[Smith and Stephenson, 2013]{Smith2013}
Smith, L.~C. and Stephenson, S.~R. (2013).
\newblock New {Trans-Arctic} shipping routes navigable by midcentury.
\newblock {\em Proceedings of the National Academy of Sciences},
  110(13):E1191--E1195.

\bibitem[Stroeve et~al., 2012]{Stroeve2012}
Stroeve, J.~C., Serreze, M.~C., Holland, M.~M., Kay, J.~E., Malanik, J., and
  Barrett, A.~P. (2012).
\newblock The {Arctic}'s rapidly shrinking sea ice cover: a research synthesis.
\newblock {\em Climatic Change}, 110(3-4):1005--1027.

\bibitem[Tietsche et~al., 2014]{Tietsche2014}
Tietsche, S., Day, J.~J., Guemas, V., Hurlin, W.~J., Keeley, S. P.~E., Matei,
  D., Msadek, R., Collins, M., and Hawkins, E. (2014).
\newblock Seasonal to interannual {Arctic }sea ice predictability in current
  global climate models.
\newblock {\em Geophysical Research Letters}, 41(3):1035--1043.

\bibitem[Wang et~al., 2013]{Wang2013}
Wang, W., Chen, M., and Kumar, A. (2013).
\newblock Seasonal prediction of {Arctic} sea ice extent from a coupled
  dynamical forecast system.
\newblock {\em Monthly Weather Review}, 141(4):1375--1394.

\bibitem[Wayand et~al., 2019]{Wayand2019}
Wayand, N.~E., Bitz, C.~M., and Blanchard-Wrigglesworth, E. (2019).
\newblock A year-round subseasonal-to-seasonal sea ice prediction portal.
\newblock {\em Geophysical Research Letters}, 46(6):3298--3307.

\bibitem[Zampieri et~al., 2018]{Zampieri2018}
Zampieri, L., Goessling, H.~F., and Jung, T. (2018).
\newblock Bright prospects for {Arctic} sea ice prediction on subseasonal time
  scales.
\newblock {\em Geophysical Research Letters}, 45(18):9731--9738.

\bibitem[Zhang and Cressie, 2019]{Zhang2019}
Zhang, B. and Cressie, N. (2019).
\newblock Estimating spatial changes over time of {Arctic Sea} ice using hidden
  2$\times$ 2 tables.
\newblock {\em Journal of Time Series Analysis}, 40(3):288--311.

\bibitem[Zhang and Cressie, 2020]{Zhang2020}
Zhang, B. and Cressie, N. (2020).
\newblock {Bayesian} inference of spatio-temporal changes of {Arctic} sea ice.
\newblock {\em Bayesian Analysis}, 15(2):605--631.

\bibitem[Zhuang, 2018]{Zhuang2018}
Zhuang, J. (2018).
\newblock xesmf: Universal regridder for geospatial data.

\bibitem[Zimmerman and Stein, 2010]{Zimmerman2010}
Zimmerman, D.~L. and Stein, M.~L. (2010).
\newblock Classical geostatistical methods.
\newblock In Gelfand, A.~E., Diggle, P., Guttorp, P., and Fuentes, M., editors,
  {\em Handbook of Spatial Statistics}, pages 29--44. CRC Press: Boca Raton,
  FL.

\end{thebibliography}

\newpage


{\Large{\bf{Appendices}}}
\begin{appendix}
\section{Standard deviation corresponding to $\gamma$ proportion of mass of a Gaussian within symmetric bounds}
 \label{app: sigmaInBounds}
Consider a Gaussian distribution with known mean $\mu = (m + M)/2$, where $M> m$. The standard deviation, $\sigma > 0$, such that $100 \times \gamma$ percent of the mass of the distribution is within $m$ and $M$ is  
\begin{equation}
\sigma = \frac{(M-m)/2}{\Phi^{-1}((1 + \gamma)/2)}
\end{equation}
where $\gamma \in (0, 1)$ and $\Phi^{-1}(\cdot)$ is the standard normal inverse cumulative distribution function.

\begin{proof}
Let $X$ be a random variable with $X \sim N(\mu, \sigma)$. Note that the value of $\sigma$ that produces  $\text{Pr}(m \leq X \leq M) = \gamma$ is equivalent to the value of $\sigma$ that produces $Pr(x \leq M) =    (1- \gamma)/2 + \gamma = (1 + \gamma)/2$. Then,
\begin{align} \nonumber
\text{Pr}\left(Z \leq \frac{M - (M + m)/2}{\sigma}\right) = \text{Pr} \left( Z \leq \frac{(M - m)/2}{\sigma} \right)  =  \frac{(1 + \gamma)}{2} 
\end{align}
Hence, 
\begin{equation}
\frac{(M - m)/2}{\sigma} = \Phi^{-1} \left(\frac{1 + \gamma}{2}\right).   \end{equation} 
\end{proof}

\section{Additional figures}
\label{app: addFigs}

\begin{figure}[!htb]
				\centering
		\includegraphics[width= .8\textwidth]{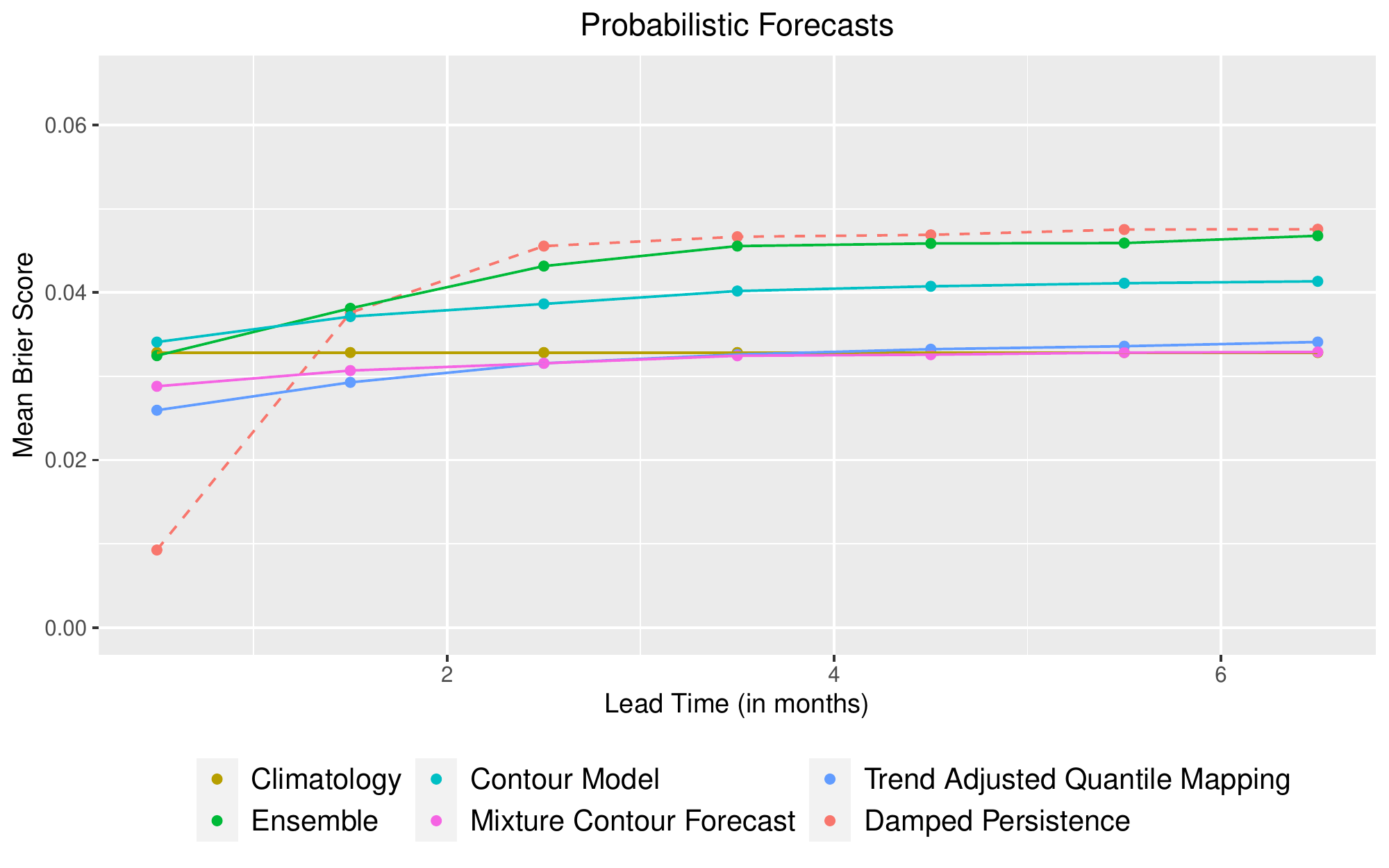} 
		\caption{Overall Brier scores  for the test years 2008-2016
for the probabilistic forecasts and a damped persistence reference binary forecast. The Brier
Score for each grid box is weighted based on its area. Forecasts are described in Table 1 in the main text.}
\label{fig: brier_prob_overall.pdf}
\end{figure}

\begin{figure}[!htb]
				\centering
		\includegraphics[width= \textwidth]{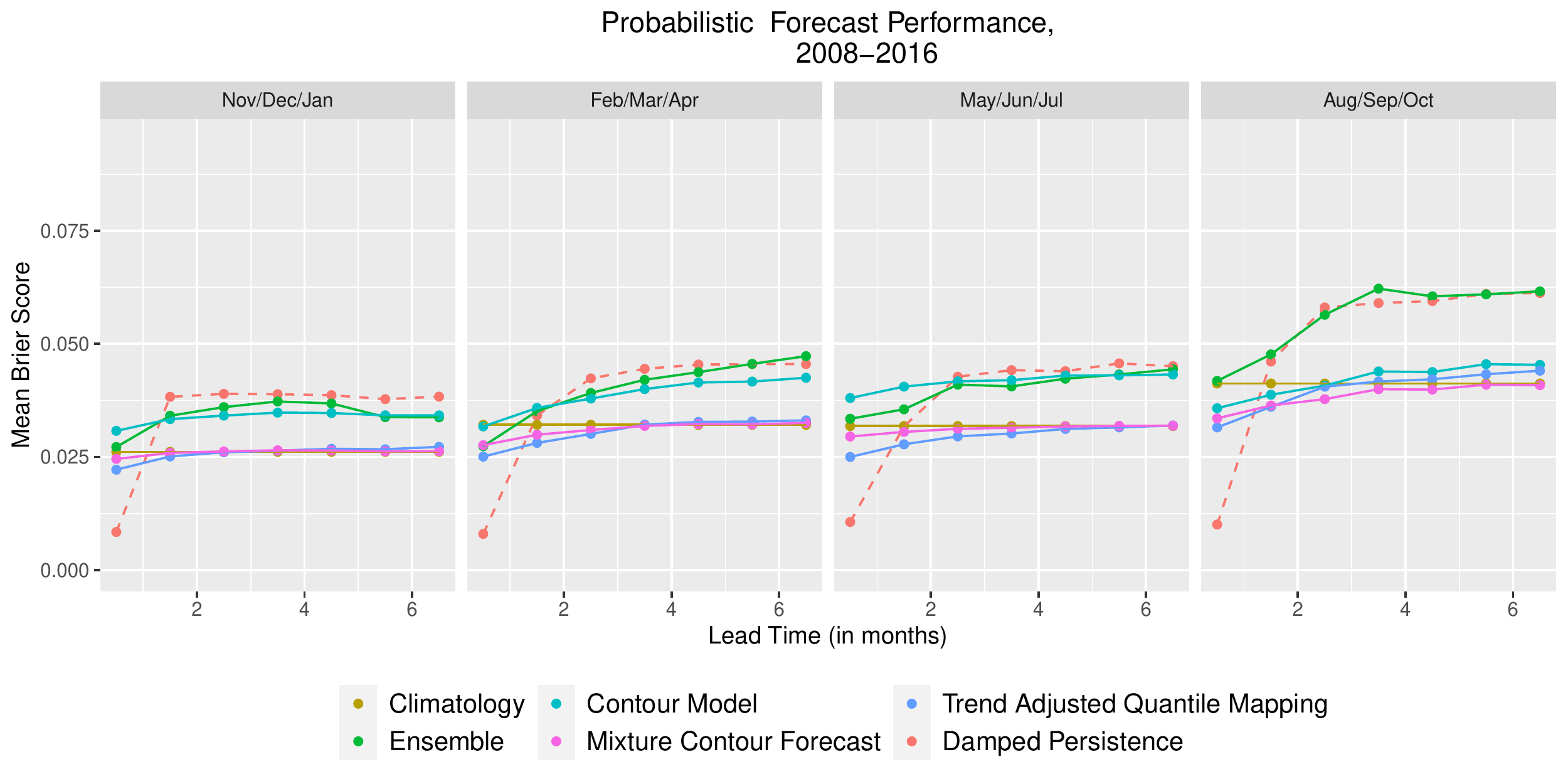} 
		\caption{Average Brier scores grouped into three-month sets for the test years 2008-2016
for the probabilistic forecasts and a damped persistence reference binary forecast. The Brier
Score for each grid box is weighted based on its area. Forecasts are described in Table 1 in the main text.}
\label{fig: brier_prob_seas}
\end{figure}

\begin{figure}[!htb]

		\centering
		\includegraphics[width= \textwidth]{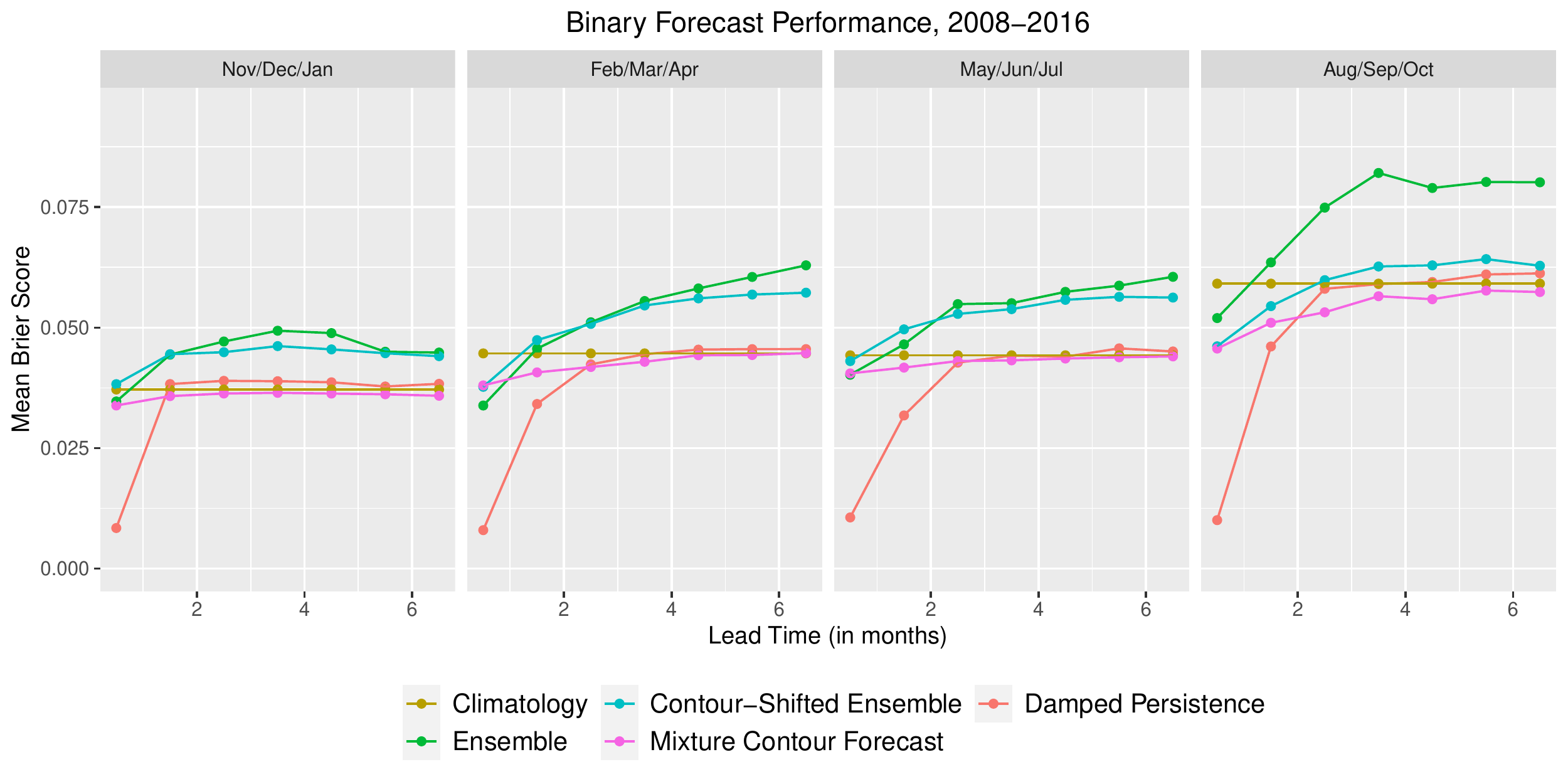} 
		\caption{As in Figure \ref{fig: brier_prob_seas}, except for binary forecasts.}
		\label{fig: calibG2}
\label{fig: brier_bin_seas}
\end{figure}

\begin{figure}[!htb]
		\centering
		\includegraphics[height  = .8\textheight]{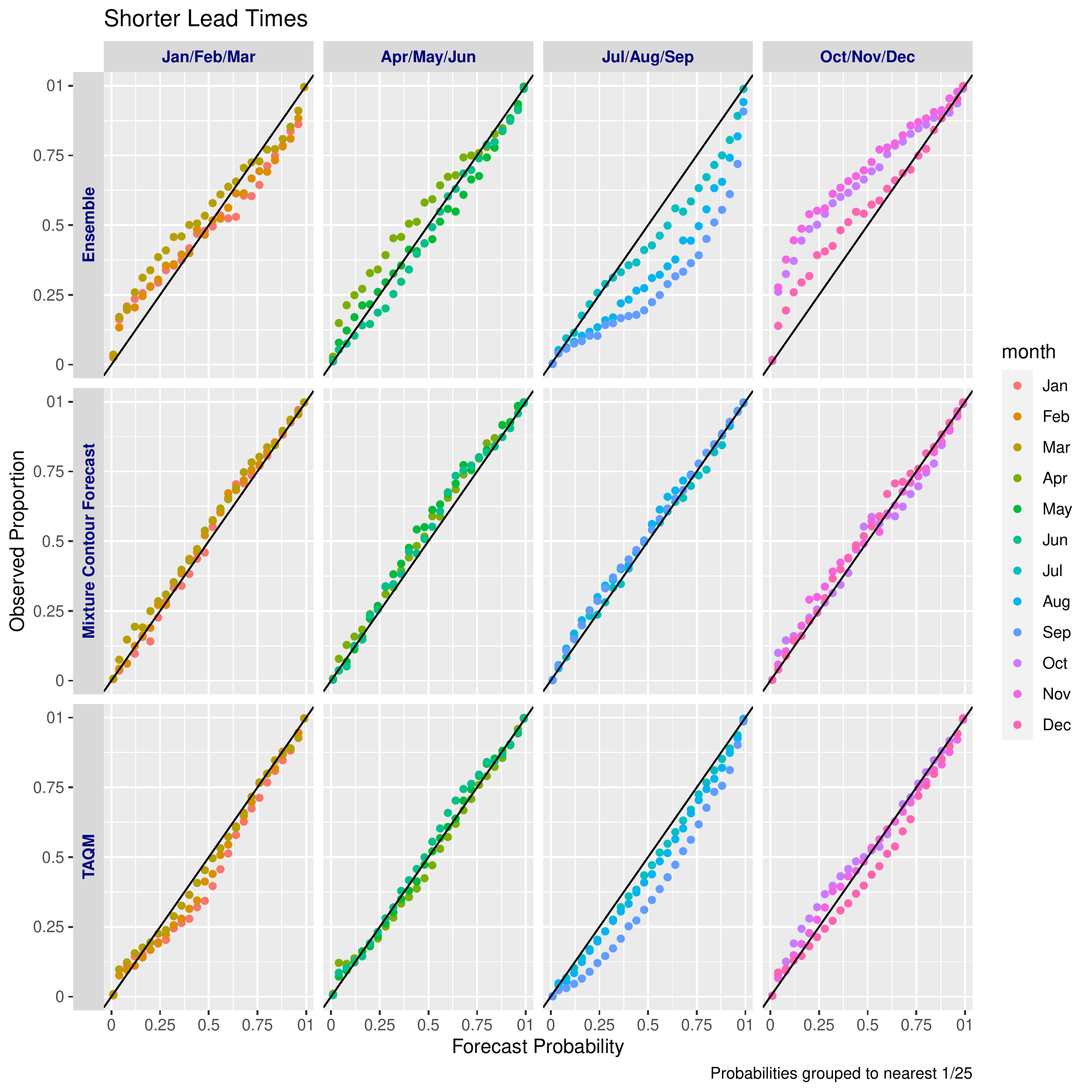} 
		\caption{Plots of the average proportion of times sea ice was present against the predicted probability of sea ice presence for the raw ECMWF (top), MCF (middle), and TAQM (bottom) forecasts for lead times of 0.5 - 1.5 months. Results are grouped into three-month sets and all grid boxes are equally weighted. }
\label{fig: calibL2}

\end{figure}

\begin{figure}[!htb]
		\centering
		\includegraphics[height  = .8\textheight]{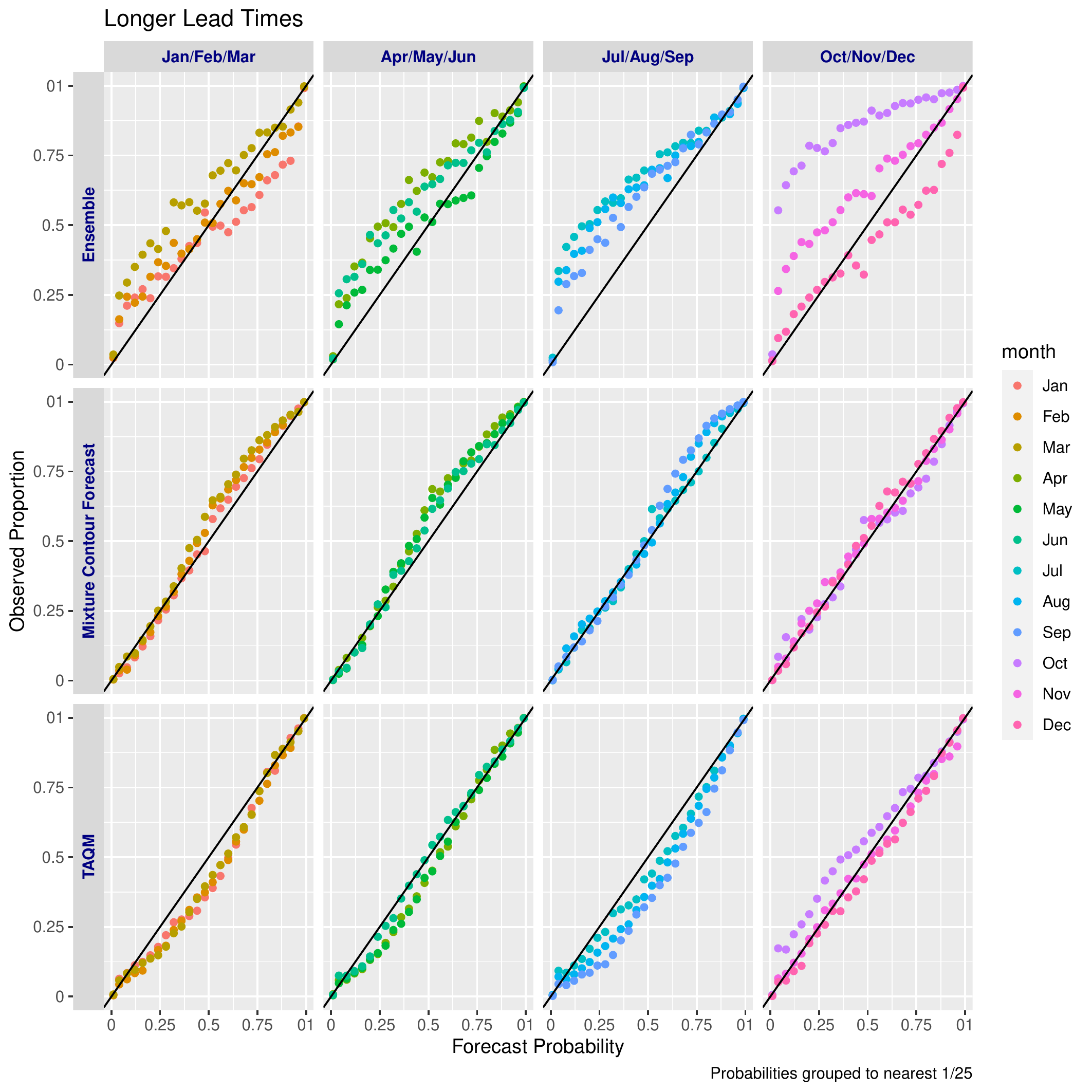} 
		\caption{As in Figure \ref{fig: calibL2} but for lead times of 2.5 - 6.5 months. }
\label{fig: calibG2}
\end{figure}

\clearpage

\section{MCMC diagnostics}
\label{app: mcmcDiag}
As in all MCMC, diagnostic analysis is needed to determine the appropriate chain length. We selected the number iterations to run the sampler by considering traceplots and model diagnostics for sample forecast months. In this section, we evaluate one month in detail. Our analysis serves a dual purpose. It demonstrates that the number of iterations used in this paper's analysis is reasonable and serves as a model for how MCMC diagnostics could be applied for future results obtained with this method. 

We note that our primary goal in this paper is prediction and not inference and that only the mean of each parameter distribution is used. So, only reasonable sampling from the posterior distribution is  needed for good performance. Therefore,  we have not repeated this analysis on every forecast month and year.  However, if our goals were to shift to inference on the parameter distributions, we would recommend doing a more thorough evaluation that would involve all parameters in each forecast month and year. We would also recommend repeating this initial diagnostic analysis if major changes are made to the method, such as changing the ensemble prediction system used in setting the hyperparameter $\boldsymbol{\mu}_{0}$. 

\subsection{Example Evaluation: September 2005, 1.5-month lead}
\noindent We evaluate the chains for September 2005 at a 1.5-month lead time using the training years of 1995-2004. We selected this month as an example, since the location of the sea ice edge is highly variable at this time of year. The model parameters are consequently likely to have high variability and need more iterations for fitting. Other months could potentially be fit adequately with less iterations.\\

 Three regions contain sea ice in September 2005 and have a contour model fit for them: the Central Arctic, Baffin Bay, and the Greenland Sea.  Using the \textit{coda} R package \citep{Plummer2006}, we compute the Raftery and Lewis Diagnostic for $\kappa$ and most values of $\mu_{i}$ and $\sigma_{i}$ \citep{Raftery1992, Raftery1995}. We use r = 0.0125 and report the maximum number of iterations needed after assessing both q = 0.025 and q = 0.975. In Table \ref{tab: mu} and Table \ref{tab: sigma}, we report the 50-th, 95-th, and 100-th  percentiles of the estimated number of iterations needed from all $\mu_{i}$ and $\sigma_{i}$ for each region. We omit from this analysis some chains for $\sigma_{i}$ and $\mu_{i}$ when more than 95$\%$ of the samples are within .05 of one of its boundaries (i.e., the upper or lower bound of the corresponding uniform prior as defined in Section 2.5 and Section 2.6 of the main text.) This omission, or something similar,  is needed because in some cases the parameter value that maximizes the posterior is on the boundary of the range. The chain will correctly not move or move little in such cases and the Raftery and Lewis Diagnostic does not make sense.  In Table \ref{tab: kappa}, we report the estimated number of iterations  needed for $\kappa$ for the three regions.
 
 \begin{table}[!htb]
\caption{The 50-th, 95-th, and 100-th percentile for the estimated chain lengths for $\mu_{i}$ obtained from the Raftery and Lewis Diagnostic for the three regions evaluated. Values rounded to the nearest whole number.   }
\label{tab: mu}
\centering
\begin{tabular}{crrr}
  \hline
 Region & $n_{est, 50}$ & $n_{est, 95}$ & $n_{est, 100}$ \\ 
  \hline
Central Arctic & 5820 & 10972 & 12715 \\ 
Baffin Bay & 3840 & 5414& 5489 \\ 
Greenland Sea & 3829 & 8704 & 8884 \\ 
   \hline
\end{tabular}
\end{table}

\begin{table}[!htb]
\caption{As in Table \ref{tab: mu}, except for $\sigma_{i}$ }
\label{tab: sigma}
\centering
\begin{tabular}{crrr}
  \hline
  Region & $n_{est, 50}$ & $n_{est, 95}$ & $n_{est, 100}$ \\ 
  \hline
Central Arctic & 5380 & 10745 & 11520 \\ 
 Baffin Bay & 5397 & 6182 & 6216 \\ 
 Greenland Sea & 7345 & 15540 & 16533 \\ 
   \hline
\end{tabular}
\end{table}

\begin{table}[!htb]
\label{tab: kappa}
\caption{Estimated chain lengths from the Raftery and Lewis Diagnostic for $\kappa$}
\label{tab: kappa}
\centering
\begin{tabular}{cr}
  \hline
 Region & $n_{\kappa}$ \\ 
  \hline
Central Arctic & 3856 \\ 
Baffin Bay & 46544 \\ 
 Greenland Sea & 36060 \\ 
   \hline
\end{tabular}
\end{table}

\begin{figure}[!htb]
		\centering
		\includegraphics[width= .8\textwidth]{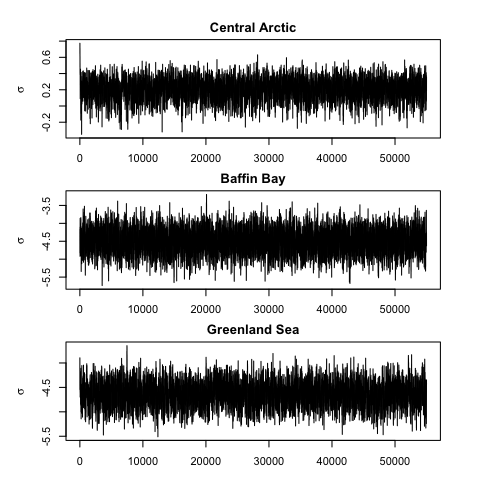} 
		\caption{Traceplots for the chains in each of the three evaluated regions for a typical $\mu_{i}$.  }
		\label{fig: tpMu}
\end{figure}

\begin{figure}[!htb]
		\centering
		\includegraphics[width= .8\textwidth]{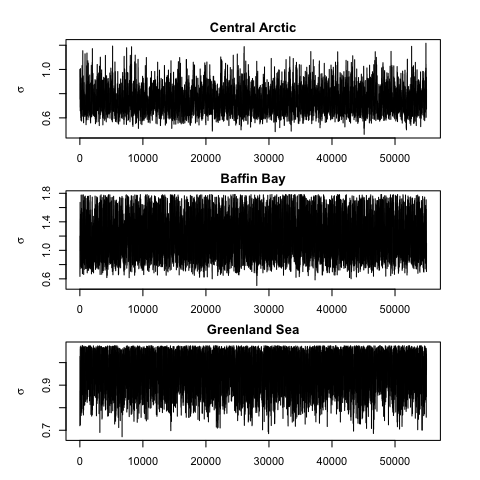} 
				\caption{Traceplots for the chains in each of the three evaluated regions for a typical $\sigma_{i}$.  }
		\label{fig: tpSigma}
\end{figure}

\begin{figure}[!htb]
		\centering
		\includegraphics[width= .8\textwidth]{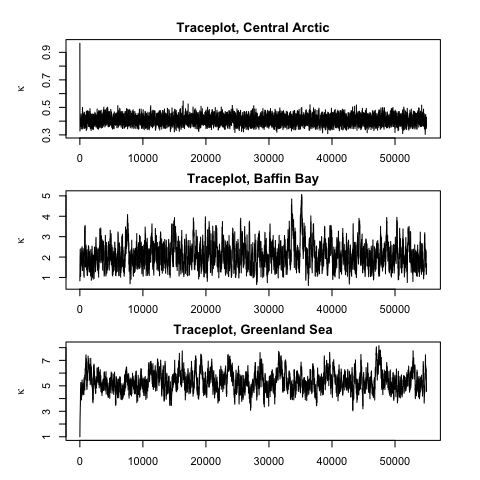} 
		\caption{Traceplots for the parameter $\kappa$ in the three evaluated regions. }
		\label{fig: tpKappa}
\end{figure}

In Figure \ref{fig: tpMu}, we show the traceplot for a typical chain for $\mu_{i}$ for each of the three regions. Figure \ref{fig: tpSigma} is an analogous figure for a typical $\sigma_{i}$. We selected the index $i$ plotted in each case by finding the chain with the estimated sample size closest to the median estimated number of iterations needed for all indices.  Finally, Figure \ref{fig: tpKappa} shows the traceplots for $\kappa$ for the three regions.

The traceplots illustrate that the chains converge quickly. The $\kappa_{i}$ parameter tends to sample the space most slowly, so $\kappa$  controls the number of iterations needed.  In particular, the maximal value from the Raftery and Lewis diagnostic is approximately 45,000. The traceplots show that a burn-in of 5000 is sufficient for the chains to have reached their posterior density region. These results motivate using 55,000 iterations for all chains in the paper with the initial 5000 iterations removed as burn-in.


\clearpage
\section{Length of Training Periods}
\label{sec: trainLengths}
Several aspects of the modeling in this paper rely on fitting parameters using multiple previous years of data as a training period. Since the Arctic is changing, more recent data is likely to be more relevant, but using only a  small number of years of data may lead to parameter estimates that are too variable. Computational cost and limited amounts of data further constrain the training lengths possible.  So, determining the appropriate training window length is not obvious.  In this section, we discuss our rationale for the values used in the paper.  

Because Contour-Shifting explicitly models the time trend, we follow \citet{Director2017} and use all available data prior to the forecast year to fit the bias correction. The earliest ensemble output ECMWF available on the Copernicus Climate Change Service Climate Data Store \citep{C3S2019} is for 1993, so that is the earliest year used in fitting Contour-Shifting.

The damped persistence forecast also relies on a trend. So, we use all available training data in fitting this reference forecast. Since this method only requires observations and they are available for earlier years than ensemble output, we fit this model with data beginning in 1981.

In the paper, we use only three years of training data in a rolling window to fit the weights in the mixture. Since the weights are dependent on results after  post-processing, there is not enough data to do a proper cross validation while leaving aside a large enough test set. We find three years works well in practice, but it is not clear this is optimal. For future use, we evaluate the performance of different training lengths for the rolling window. We fit the weights and corresponding MCF forecasts for the years 2012-2016 using rolling windows of training lengths from 1-7 years. We report the mean Brier score over months and years. As in the main paper, we weight the grid boxes by their area. We find that a five-year training period performs best and would recommend this training length for operational use. We do note however that forecast performance does not appear to be particularly sensitive to this choice.

\begin{table}[ht]
\caption{Mean area-weighted Brier scores for MCF on the test set of 2012-2016 for different numbers of years of training data used to determine the weight on the climatology versus the contour model forecast.}
\centering
\begin{tabular}{rlr}
  \hline
  Years &Mean Brier Score \\ 
  \hline
  1 & 0.03313 \\ 
  2 & 0.03317 \\ 
    3 & 0.03315 \\ 
     4 & 0.03293 \\ 
   5 & 0.03281 \\ 
    6 & 0.03286 \\ 
  7 & 0.03293 \\ 
   \hline
\end{tabular}
\end{table}

\end{appendix}
\end{document}